\documentclass[fleqn,usenatbib]{mnras}
\usepackage{newtxtext,newtxmath}
\usepackage[T1]{fontenc}

\DeclareRobustCommand{\VAN}[3]{#2}
\let\VANthebibliography\thebibliography
\def\thebibliography{\DeclareRobustCommand{\VAN}[3]{##3}\VANthebibliography}

\usepackage{graphicx}	% Including figure files
\newcommand{\kms}{{km\,s}$^{-1}$}
\newcommand{\teff}{$T_\mathrm{eff}$\,}
\newcommand{\logg}{$\log g$\,}

\newcommand{\Msun}{\,$\rm{M}_\odot$}

\newcommand{\vsini}{$v_{\rm{e}} \sin i$}
\newcommand{\vsinif}{$v_{\rm{e}} \sin i_{\rm{F}}$}
\newcommand{\ve}{$v_{\rm{e}}$}
\newcommand{\vorb}{$v_{\rm{orb}}$}
\newcommand{\vlc}{$v_{\rm{c}}$}
\newcommand{\vc}{$v_{\rm{crit}}$}
\newcommand{\vr}{$v_{\rm{r}}$}
\newcommand{\Dvr}{$\Delta v_{\rm{r}}$}

\newcommand{\nc}{$N_{\rm{crit}}$}

\newcommand{\fc}{$F_{\rm{crit}}$}

\newcommand{\gae}{$\gamma_{\rm{e}}$}
\newcommand{\gmax}{$\gamma_{\rm{max}}$}
\newcommand{\gc}{$\gamma_{\rm{cent}}$}

\title[Be-type stars in 30 Dor]{Properties of the Be-type stars in 30 Doradus}

\author[P. L. Dufton et al.]{P. L. Dufton,$^1$ D. J. Lennon,$^2$ J. I. Villase\~nor,$^3$ I.D. Howarth,$^4$ C. J. Evans,$^5$ S. E. de Mink,$^6$ H. Sana$^7$ \newauthor and W. D. Taylor$^5$
\\
$^1$Astrophysics Research Centre, School of Mathematics and Physics, Queen's University Belfast, Belfast BT7 1NN, UK\\
$^2$Instituto de Astrof\'isica de Canarias, E-38200 La Laguna, Tenerife, Spain\\
$^3$Institute for Astronomy, University of Edinburgh, Royal Observatory, Blackford Hill, Edinburgh, EH9 3HJ, UK \\
$^4$Department of Physics and Astronomy, University College London, Gower Street, London, WC1E 6BT, UK\\
$^5$UK Astronomy Technology Centre, Royal Observatory Edinburgh, Blackford Hill, Edinburgh, EH9 3HJ, UK\\
$^6$Max-Planck-Institut f\"ur Astrophysik, Karl-Schwarzschild-Strasse 1, 85741 Garching, Germany\\
$^7$Institut voor Sterrenkunde, Universiteit Leuven, Celestijnenlaan 200 D, B-3001 Leuven, Belgium
}

\date{Accepted XXX. Received YYY; in original form ZZZ}

\pubyear{2021}

\hypersetup{draft}

\begin{document}
\label{firstpage}
\pagerange{\pageref{firstpage}--\pageref{lastpage}}
\maketitle

\begin{abstract}
The evolutionary status of Be-type stars remains unclear, with both single-star and binary pathways having been proposed. Here, VFTS spectroscopy of 73 Be-type stars, in the spectral-type range, B0--B3, is analysed to estimate projected rotational velocities, radial velocities and stellar parameters. They are found to be rotating faster than the corresponding VFTS B-type sample but simulations imply that their projected rotational velocities are inconsistent with them all rotating at near critical velocities. The de-convolution of the projected rotational velocities estimates leads to a mean rotational velocity estimate of 320--350\,\kms, approximately 100\,\kms\ larger than that for the corresponding B-type sample. There is a dearth of targets with rotational velocities less than 0.4 of the critical velocity, with a broad distribution reaching up to critical rotation. Our best estimate for the mean or median of the rotational velocitiy is 0.68 of the critical velocity. Rapidly-rotating B-type stars are more numerous than their Be-type counterparts, whilst the observed frequency of Be-type stars identified as binary systems is significantly lower than that for normal B-type stars, consistent with their respective radial-velocity dispersions. The semi-amplitudes for the Be-type binaries are also smaller. Similar results are found for a SMC Be-type sample centred on NGC\,346 with no significant differences being found between the two samples. These results are compared with the predictions of single and binary stellar evolutionary models for Be-type stars. Assuming that a single mechanism dominated the production of classical Be-type stars, our comparison would favour a binary evolutionary history. 
\end{abstract}


%%%%%%%%%%%%%%%%%%%%%%%%%%%%%%%%%%%%%%%%%%%%%%%%%%

%%%%%%%%%%%%%%%%% BODY OF PAPER %%%%%%%%%%%%%%%%%%

%________________________________________________________________
\section{Introduction}\label{s_intro}
An important subgroup within the B-type stellar population are Be-type stars, which exhibit strong emission in their Balmer lines \citep{col87,jas90}. Generally the Be-type classification excludes supergiants, young Herbig stars, mass transfer binaries and B[e] type stars as discussed by, for example, \citet{riv12a}. \citet{str31} first identified that Be-type stars rotated relatively rapidly and that their emission lines could be ascribed to a circumstellar disc. In our Galaxy, \citet{zor97} found approximately 17\% of B-type stars to be Be-type, with estimates of 17-20\% in the Large Magellanic Cloud \citep{mar06a} and 26-40\% in the Small Magellanic Cloud \citep{mar07a}. \citet{mart07} suggested that the higher percentage for the SMC might be due to systematic larger stellar rotational velocities in this lower metallicity environment.

The rotational velocities of Be-type stars have been studied by many authors with the earlier investigations being critically reviewed by \citet{riv12a}. They concluded that Be-type stars typically rotated at a fraction of approximately 0.8 of the Keplerian circular orbital velocity. Subsequently \citet{zor16} studied the rotational properties of a large sample of Galactic Be-type stars. They deconvolved the projected rotational velocities (\vsini) estimates to deduce the underlying distribution of equatorial radial velocities. After carefully considering possible systematic errors in both the \vsini-estimates and stellar parameters, they concluded that the mode of the equatorial rotational velocity as a fraction of the critical velocity was approximately 0.65. Recently \citet{bal20, bal21} have discussed TESS photometry of over 400 Galactic Be-type stars and deduced their rotational periods and hence equatorial rotational velocities. Their results are in excellent agreement with those of \citet{zor16}. These more recent results are difficult to reconcile with all Be-type stars being near critical rotators and indeed \citet{zor16} commented that `the probability that Be stars are critical rotators is extremely low' 

The current evolutionary status of Be-type stars remains unclear, with theoretical models falling into two broad categories, viz.\ those that postulate that Be-type stars have evolved as single stars or that they are the product of previous mass transfer in a binary system. For the former (which could apply to both single and non-interacting binary systems), the Be-type properties would simply arise from their large rotational velocities relative to their critical velocities. This could be due to high primordial rotational velocities although as discussed by \cite{bod20} this would appear to be inconsistent with observational studies of stellar rotation rates. Alternatively it could arise from changes in the rotational velocity and the critical velocity as a B-type star evolves from the zero age main sequence. Recently \citet{has20} have discussed this scenario and find that equatorial velocities that are near to critical should occur. However their comparison with the observed properties of Be-type stars highlighted a number of inconsistencies. Additionally their modelling assumed that all rapidly rotating objects would be Be-type stars, whilst as discussed in Sect.\ \ref{d_fast} this is not the case.

Evolutionary histories involving binary interactions have also been extensively explored \citep[see, for example,][]{pol91, deM13, riv12a, sha14, sha20, has21}. In this scenario the observed Be-type star was originally the less massive component and has accreted mass and angular momentum from the more massive primary, which if it does not become a supernova, can evolve into a low mass helium star or white dwarf. 

The VLT-FLAMES Tarantula Survey (VFTS) obtained multi-epoch optical spectroscopy of over 800 massive stars in the 30 Doradus region \citep{eva11} of which 438 were classified as B-type, including 73 Be-type \citep{eva15}. The only selection criterion  was a magnitude cut-off at V\,=\,17, in order to provide sufficient signal-to-noise ratios in the spectroscopy. As such it provides a relatively unbiased Be-type sample, together with a comparable B-type sample. Because of the magnitude cut-off, both samples are limited to the earlier B-types, viz B0--B3. Further details of the samples and their biases can be found  as supplementary material in Appendix E. Below we analyse these samples to investigate rotational and radial velocities, stellar parameters, and binarity. We also consider comparable samples of Be-type and B-type stars in NGC\,346 in the Small Magellanic Cloud \citep{duf19}. These results are used to address several issues for Be-type stars and in particular their evolutionary status and whether they are all rotating at near critical velocities.

%________________________________________________________________
\section{Observations}\label{s_obs}

\subsection{Target selection and data reduction} \label{s_red}

\begin{figure*}
	\includegraphics[width=19cm, angle=0]{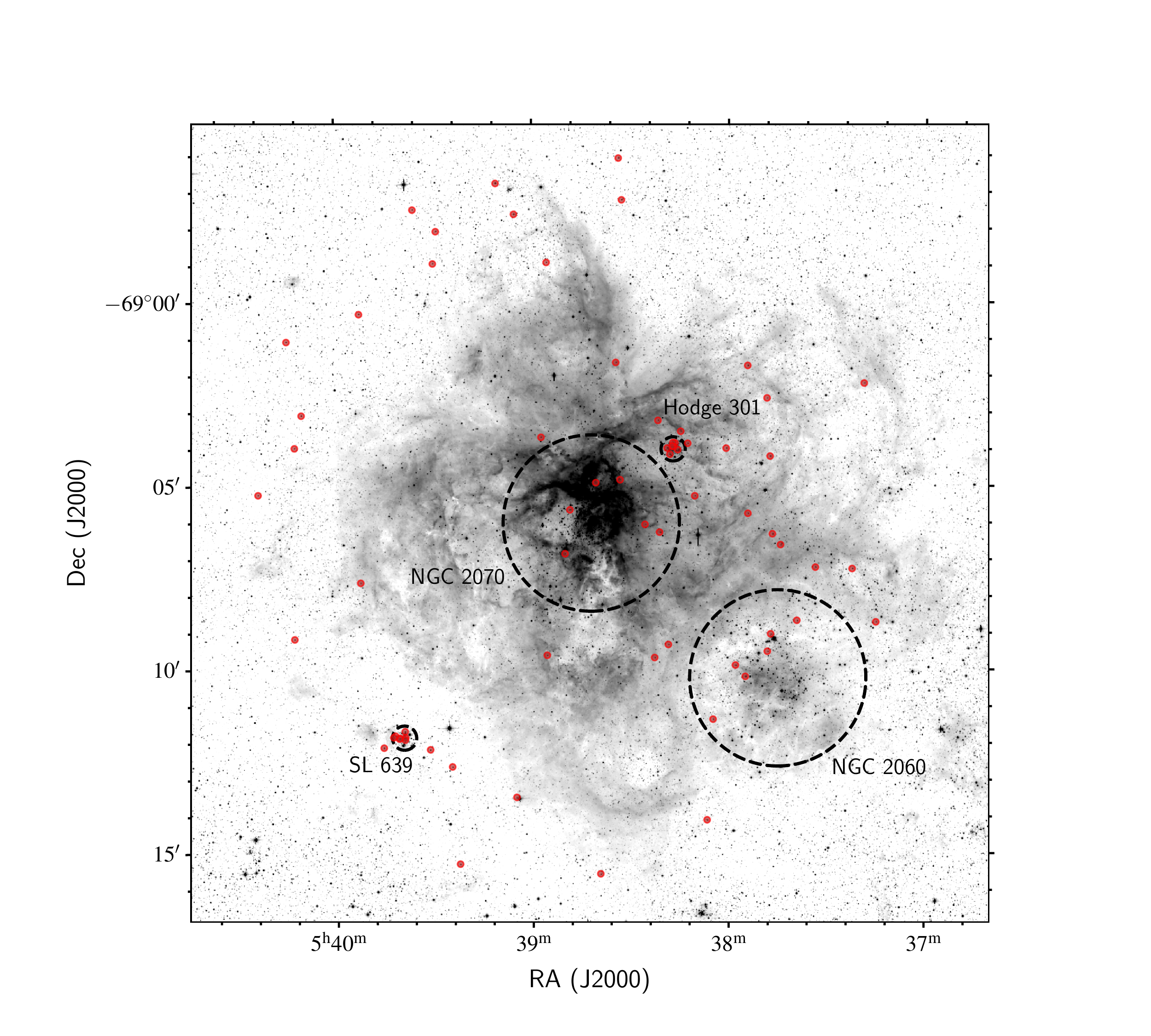} 	
	\caption{Spatial distribution of the Be-type targets in 30 Doradus. The spatial extents of NGC 2070, NGC 2060, SL 639, and Hodge 301 \citep{eva15} are indicated by the overlaid dashed circles. The underlying image is from a V-band mosaic taken with the ESO Wide Field Imager on the 2.2 m telescope at La Silla.}
	\label{f_spatial}
\end{figure*}

The VFTS spectroscopy was obtained using the MEDUSA mode of the FLAMES instrument \citep{pas02} on the ESO Very Large Telescope and has been discussed in detail by \citet{eva11}. MEDUSA uses fibres to simultaneously `feed'  the light from over 130 sky positions to the Giraffe spectro\-graph. Nine fibre configurations (designated fields~`A' to `I' with near-identical field centres) were observed in the 30~Doradus region, sampling different clusters and the local field population. Three standard Giraffe grating settings were used viz.\ LR02 (wavelength range from 3960 to 4564\,\AA\ at a spectral resolving power of R$\sim$7\,000), LR03 (4499--5071\,\AA, R$\sim$8\,5000) and HR15N (6442--6817\AA, R$\sim$16\,000). Spectroscopy of more than 800 early-type stars was obtained with 73 targets subsequently being classified as Be-type by \citet{eva15}, from the identification of stellar H$\alpha$\ emission in HR15N spectroscopy. Details of the target selection, observations, and initial data reduction have been given in \citet{eva11},  where target co-ordinates were also provided. Fig.\ \ref{f_spatial} illustrates the spatial positions of all the Be-type stars that have been analysed.

For the analyses presented, both the LR02 and LR03 spectroscopy was utilised with all useable exposures of a given target being combined (after normalisation) using either a median or weighted $\sigma$-clipping algorithm. The final spectra from the different methods were normally indistinguishable.  The combined spectra for some features (e.g. the Balmer series) were subsequently re-normalised to ensure consistency with the normalisations adopted in the theoretical predictions. 

\citet{gar17} have discussed the estimation of projected rotational velocities (\vsini) for SB1 candidates from spectroscopy that had been combined without wavelength shifts. They concluded that this would not lead to significant errors provided the range of radial velocities (\Dvr) was relatively small. For stars with negligible rotation, the limit was \Dvr$\la$30\,\kms, increasing to, for example, 120\,\kms\ for \vsini$\sim$200\,\kms. All the SB1 candidates considered here fulfil these criteria and therefore no velocity shifts have been applied to exposures at different epochs.

We have searched the Hubble Legacy Archive to search for spatially close companions, with imaging available for over 80\% of our sample. Nine targets had possible contamination, with that for four targets (\#272, \#283, \#644, and \#824) being potentially serious. Further details can be found in the supplementary material in Appendix A.

\subsection{Binarity} \label{s_bin}

\citet{dun15} identified seven of our targets as SB1 candidates and all apart from \#135 have been further investigated by \citet{vil21}. For \#135,  \citet{dun15} considered two \ion{He}{i} lines at 4026 and 4387\AA\ and their radial velocity estimates at a given epoch differed by 20-60\,\kms, probably due to the very strong nebular emission for these lines. We have repeated their cross-correlation analysis but now for three weaker \ion{He}{i} lines at 4009, 4121 and 4144\AA. Inspection of these features shows little nebular emission, whilst the profiles appear symmetric. We obtain a smaller range in radial velocity estimates between epochs of 45\,\kms with typical uncertainties of $\pm$15\,\kms. As such these variations do not fulfil one of the criteria of \citet{dun15} for binarity, viz. that the range should be more than four times the estimated uncertainties. Hence this target has been removed from the SB1 candidates. The status of the other 6 SB1 candidates is discussed in Sect.\ \ref{d_bin}.

%________________________________________________________________
\section{Projected rotational velocity estimates}\label{s_vsini}

\subsection{Methodology}\label{s_vini_meth}

Most of our targets have published estimates for their projected rotational velocity, \vsini\  \citep{duf12}. However, given the complexities of Be-type spectra, our approach has been to re-estimate these and also to attempt to obtain estimates for the targets which had not been previously considered. We have used a Fourier Transform (FT) methodology \citep{car33, sim07} to estimate the projected rotational velocities following the procedures described in \citet{duf12}. This approach has been widely used for early-type stars \citep[see, for example][]{dufsmc06, lef07, mar07, sim10, fra10, duf12, sim14, sim17} and relies on the convolution theorem \citep{gra05}, viz. that the Fourier transform of convolved functions is proportional to the product of their individual Fourier Transforms. It then identifies the first minimum in the Fourier transform for a spectral line, which is assumed to be the first zero in the Fourier transform of the rotational broadening profile with the other broadening mechanisms exhibiting either no minima or only minima at higher frequencies. Further details on the implementation of this methodology are given by \citet{sim07} and \citet{duf12}. 

As discussed by \citet{sim07}, the first minimum in the Fourier Transform can be difficult to identify in spectra with, for example, low SNRs or significant nebular contamination.  Therefore we have also fitted rotational profiles to the observed spectral features. As we have not included either the intrinsic or instrumental broadening, the corresponding \vsini\ estimates may be overestimated, especially when the projected rotational velocity is relatively small -- further details can be found in \citet{duf12, ram13, ram15}. However these profile fitting (PF) estimates still provide a useful check on the FT estimates. 

Two different sets of spectral lines have been adopted for analysis, depending on the degree of rotational broadening, and they are summarized in Table \ref{t_lines}. For the narrower-lined stars (with \vsini$\la 150$\kms), a selection of metal lines and non-diffuse helium lines  was used. These have the advantage of being intrinsically narrow, whilst the former are not normally affected by nebula emission. For larger rotational velocities, reliable results could only be obtained from the stronger absorption lines and in these cases a selection of diffuse and non-diffuse helium lines was considered. Some features are close doublets or triplets, whilst additionally the \ion{He}{i} diffuse lines have weak forbidden components in their blue wing.\footnote{For the spectral types considered here, blending of the \ion{He}{i} line at 4026\AA\ with a \ion{He}{ii} line should be unimportant.}  For the former, the separations were less than the broadening due to rotation and indeed the instrumental profile, whilst the latter were only used for the more rapidly rotating stars where the profiles appeared symmetric within the observational uncertainties.

\begin{table}
	\caption{Absorption lines (and their approximate wavelength in \AA) that were used in estimating the projected rotational velocity. Possible blends are noted and for the \ion{He}{i} features, those particularly prone to contamination by nebular emission are identified.}\label{t_lines}
	\begin{center}
		\begin{tabular}{llrlccrccccc} 
			\hline\hline
			
			Species       &  $\lambda$  &   Set &  Comment                        \\
			\ion{He}{i}   &   4009	    &	2	&                                 \\
			\ion{He}{i}   &   4026	    &	2	&  Nebular emission               \\
			\ion{He}{i}   &   4121	    &	1,2	&  \ion{O}{ii} line at 4419\AA    \\
			\ion{He}{i}   &   4144	    &	2	&                                 \\
			\ion{He}{i}   &   4169	    &	1	&                                 \\
			\ion{C}{ii}   &   4267	    &	1	&                                 \\
			\ion{He}{i}   &   4387	    &	2	&                                 \\
			\ion{He}{i}   &   4471	    &	2	&  Nebular emission               \\
			\ion{Mg}{ii}  &   4481	    &	1	&  \ion{Al}{iii} line at 4479\AA  \\
			\ion{Si}{iii} &   4553	    &	1	&                                 \\
			\ion{Si}{iii} &   4568	    &	1	&                                 \\
			\ion{He}{i}   &   4713	    &	1,2	&                                 \\
			\ion{He}{i}   &   4922	    &	2	&                                 \\			
			\hline
		\end{tabular}
	\end{center}
\end{table}

Of the 73 VFTS Be-type targets \citep{eva15}, previous projected rotational velocities were available for 64 \citep{duf12, gar17}. New estimates have been obtained here for all these targets plus five additional targets (\#022, \#030, \#272, \#293, and \#395). The spectroscopy of the remaining 4 targets was not analysed as discussed in the supplementary material in Appendix A.

We have assigned a quality flag, Q, to our estimates for each target as specified below:

\noindent 
\underline{Class 1}: Well observed \ion{He}{i} or metal line spectrum with estimates being obtained from most lines. Features have a bell-shaped profile (supplemented by more extended wings for diffuse \ion{He}{i} transitions, which exhibit linear Stark effect broadening), which are well fitted by a rotationally broadened profile with the appropriate \vsini-estimate. Little or no evidence for shell-like absorption or emission features. Some profiles may be affected by narrow emission (or absorption) features due to uncertainties in the removal of the nebular emission. 

\noindent 
\underline{Class 2:} As for Class 1 but the intrinsic weakness of the spectral features (often coupled with a moderate S/N) limits the number and accuracy of the \vsini-estimates. Additionally some \ion{He}{i} lines (normally triplets) may show significant nebular contamination.

\noindent 
\underline{Class 3:} Problematic spectra discussed in the supplementary material in Appendix A.

\begin{table*}
	\caption{Estimates of the projected rotational and radial velocities. For the former, \vsini\ is the weighted mean of the Fourier-Transform  estimates, $\sigma$ the standard deviation, $n$ the number of estimates, Q the quality estimate, whilst PF is the mean of the profile-fitting estimates. Also listed are estimates of the upper limits of the projected rotational velocity (\vsinif) after applying the  corrections discussed in Sect.\ \ref{s_vsini_bias}. Spectral types and sub-cluster membership (H: Hodge\,301, S: SLS\,639) are taken from \citet{eva15}, whilst systems identified as SB1 binaries by \citet{dun15, vil21} are marked by asterisks. As discussed by \citet{eva06}, the "Be+" classification is on based on the identification of \ion{Fe}{ii} emission features. The full table is available as supplementary material.}\label{t_vsini}
	\begin{center}
		\begin{tabular}{lllr@{\hskip 0.4in}cccrccc@{\hskip 0.4in}cccc} 
			\hline\hline			
			VFTS	&	Spectral Type	& Cluster& S/N	 &	\multicolumn{7}{c}{Projected rotational velocity~~~~~~~~~~}	&	\multicolumn{2}{c}{\!\!\!\!\!Radial velocity} \\
			&                           &        &       & Set	 &	\vsini	&	$\sigma$	&	$n$	& Q & PF  & \vsinif	&~~~\vr & $\sigma$
			\\
			022     &   B0-0.5 V-IIIe   &    --  &   65  &   2   &   364 &   36  &   4  &   3   &   312  &   406 &   262 &   26  \\
			030     &   B3-5e (shell)   &    --  &   65  &   2   &   295 &   20  &   4  &   2   &   309  &   317 &   263 &   17  \\
			034     &   B1.5 Ve         &    --  &   105 &   2   &   174 &   15  &   8  &   1   &   199  &   190 &   306 &   20  \\
			068     &   B1-1.5e         &    --  &   55  &   2   &   337 &   66  &   3  &   3   &   348  &   370 &   263 &   26  \\
			101     &   B0.7: Vne       &    --  &   95  &   2   &   358 &   19  &   7  &   1   &   342  &   398 &   273 &   21  \\
  			\hline
\end{tabular}
\end{center}
\end{table*}

Means of the \vsini~FT estimates for each star have been calculated by weighting individual estimates by the central depth of the line, to approximately allow for the differences in the observational quality. Unweighted means showed no systematic offset and agreed to typically 3\,\kms with a maximum difference of 10\,\kms. Table \ref{t_vsini} summarizes weighted means, their standard deviations, the number of lines considered and their quality rating, Q. Also listed are spectral types \citep{eva15}, \vsini-estimates from the PF methodology and signal-to-noise ratios (S/N). The last have been estimated for the wavelength region, 4200-4250\,\AA, which should not contain strong absorption lines. However particularly for the higher estimates, these should be considered as lower limits as they could be affected by weak features. S/Ns for the LR03 region were normally similar or slightly smaller \citep[see, for example,][for a comparison of the ratios in the two spectral region]{mce14}. 

The fits for the Class 1 objects were generally convincing and the means of the \vsini-estimates were normally based on at least 6 independent measurements. Exceptions were \#233 and \#876 where only 5 estimates were available due to a relative low S/N in the LR03 spectroscopy and \#874 where again 5 estimates were available due to the metal line spectrum being relatively weak. However the fits for these three targets appear reasonable whilst the standard deviation of the individual estimates are consistent with those for other targets with the highest quality rating. 

For most of the Class 2 objects, the agreement between observation and the theoretical profiles was also reasonable, although now the number of estimates was generally smaller and/or nebular emission was present in some lines. In Appendix A (available as supplementary material), we comment on two Class 2 targets where the spectra presented additional problems. The estimates for the 14 targets with the lowest quality rating were affected by a variety of problems including strong nebular emission, low S/N spectroscopy and close companions and are all discussed in Appendix A.

In Appendix B (available as supplementary material), we discuss the stochastic uncertainties in the \vsini-estimates, including those obtained from the different methodologies and different line-sets. We also compare our estimates with those obtained previously and conclude that a typical stochastic uncertainty would be 3-4\%, apart from the targets with Q=3, where it would be larger.

\subsection{Systematic errors in \vsini-estimates} \label{s_vsini_bias}

For rapidly rotating stars, it is also important to consider systematic biases in the \vsini-estimates \citep{tow04, cra05, fre05, zor16}.  Their magnitude will depend on the stellar parameters some of which (for example, the angular velocity) may be unknown. Our approach has been to try to identify the largest plausible corrections, thereby estimating upper limits for the \vsini-distribution and in turn for the rotational velocity distribution.

The effects of stellar rotation on early-type stellar spectra have been investigated for over 40 years \citep[see, for example,][and references therein]{col77, col91, tak17, abd20}. Here we utilise the simulations of \citet{tow04} and \citet{fre05} for B-type stars, including the early to mid-B-type stars, present in our samples.
\citet{tow04} considered spectral types between B0 and B9 and discussed possible biases for two lines considered here (\ion{He}{i} 4471\AA\ and \ion{Mg}{ii} 4481\AA). The \vsini-underestimation was largest for targets at a near critical rotational velocity and observed at an angle of inclination, $i\sim 90^{\circ}$. Our targets have spectral types in the range B0 to B3. For the former, they found underestimates of up to 12\% (\ion{He}{i} 4471\AA) and 9\% (\ion{Mg}{ii} 4481\AA), increasing to up to 22\% and 12\% respectively for the B3 spectral type. To estimate the biases in our results, we have considered a spectral type B2 and used their results for the \ion{He}{i} line as this ion was used for the broader lined targets and had the larger corrections. From their simulations, we can find the maximum correction as a function of the ratio of the measured \vsini\ to the critical velocity (\vc). This correction was  negligible for ratios of less than one half, increasing to reach more than 20\% at critical rotation. 

\citet{fre05} considered the same two lines and provided corrections for 3 effective temperatures (15\,000, 20\,000 and 24\,000K) and 3 gravities (3.4, 3.8 and 4.2 dex). In agreement with \citet{tow04}, they found that these were larger for helium than for magnesium, that they were anti-correlated with effective temperature, and that they fell very rapidly with decreasing angular velocity. We have adopted their corrections for \teff=24\,000K, and \logg= 3.8\,dex in good agreement with the median atmospheric parameters of our dataset (see. Sect.\ \ref{s_par}). Additionally in order to estimate the maximum corrections, we used their simulations for an angular velocity 99\% of critical. For both sets of corrections, we fitted a cubic polynomials with the appropriate independent variable: apparent \vsini\ for \citet{fre05} and the ratio of the apparent \vsini\ to the critical velocity for \citet{tow04}. Corrections derived from these formulae agreed to typically $\pm1$\,\kms with the original values. 

For our sample, the correction deduced from the two sets of simulations\footnote{Those for \citet{tow04} used the critical velocities in Table \ref{t_atm}} were qualitatively similar but those from \citet{fre05} were systematically larger. For example, the mean difference in the corrections was 13\,\kms, whilst that for the top quartile of \vsini-estimates was 16\,\kms. These differences are not surprising as the corrections are based on different assumptions. In particular those of \citet{fre05} can be considered as upper limits as we have assumed that all targets are rotating at effectively the critical velocity. In Table \ref{t_vsini}, projected rotational velocity estimates (\vsinif) including these corrections are listed and will be discussed in Sect.\ \ref{s_disc}.

%________________________________________________________________
\section{Radial velocity estimates}\label{s_vr}

Mean stellar radial velocity estimates have previously been estimated by \citet{eva15} for most of our targets. They were based on fitting Gaussian profiles to observed \ion{He}{i} and/or metal lines. Given their relatively large projected rotational velocity, we have re-determined radial velocities for all the 63 targets which were not identified as SB1 systems. We utilised the central wavelengths found from profile fitting as discussed in Sect.\ \ref{s_vsini}, thereby limiting our estimates to those lines that had been used to estimate projected rotational velocities. 

Initially we adopted laboratory wavelengths taken from the NIST Atomic Spectra Database \citep{NIST_ASD}. However all the diffuse helium lines contain a forbidden (P-F) component in their blue wing, together with other smaller asymmetries \citep{gri97}. Evidence for this was found in the systematic offsets between the radial velocity estimates from different \ion{He}{i} lines. In particularly the largest negative offsets were found for the diffuse triplet (at 4026 and 4471\AA), where the forbidden components are strongest.

We have therefore undertaken simulations to quantify the effects of these asymmetries for line-set 2 (see Table \ref{t_lines}); no simulations were undertaken for line-set 1 as these should be symmetric. Three effective temperatures (20000, 23000 and 27500\,K) and three gravities (3.0, 3.5 and 4.0 dex) were considered leading to a total of nine combinations. This grid covered the estimated atmospheric parameters of over 80\% of the sample (see Table \ref{t_atm}), with the remainder lying close to its edges. For each set of atmospheric parameters, helium line profiles were taken from the models discussed in Sect.\ \ref{s_par} and convolved with a rotational broadening function for \vsini-values of 175, 275 and 400\,\kms. Each profile was then fitted with a rotational broadening profile following the procedure used when analysing the observational spectroscopy. For each line and choice of \vsini, we calculated the mean of the nine central wavelength estimates and their standard deviations. These are summarized in Table \ref{t_vr_wave} together with the means and standard deviations averaged over all the simulations for a given line. For the non-diffuse (4121 and 4713\AA) and the singlet diffuse (4009, 4143, 4387 and 4921\AA) features, the mean wavelengths are in excellent agreement with the laboratory values. This is consistent with these features having theoretical profiles with a high degree of symmetry. By contrast the simulations for the two diffuse triplets (4026 and 4471\AA) give central wavelengths that are smaller by $\sim$0.2--0.3\AA\ than the laboratory wavelengths of the allowed components. These shifts would be consistent with the observational offsets found from these lines when adopting laboratory wavelengths (see supplementary material in Appendix C).

Adopting the wavelengths implied by the simulations (last column of Table \ref{t_vr_wave}), we summarize the mean radial velocity estimates and their sample standard deviations in Table \ref{t_vsini}. These uncertainties are discussed in Appendix C (available as supplementary material), whilst estimates have not been undertaken for the  targets classified as SB1 in Sect.\ \ref{s_obs}. The same features were used as in the projected rotational velocity estimation and therefore the number of radial velocity estimates is again given by the values of $n$\ listed in Table \ref{t_vsini}.

\begin{table*}
	\caption{Laboratory and simulated wavelength for the \ion{He}{i} lines used in the estimation of the radial velocity.}\label{t_vr_wave}
	\begin{center}
		\begin{tabular}{llccccccccccc} 
			\hline\hline			
			Line  & Laboratory &   \multicolumn{3}{c}{Projected rotational velocity} & All\\
			&            &   175\kms
			&   275\kms
			&   275\kms\\
			
			\hline                   
			4009 & 4009.26 & 4009.24$\pm$0.02 & 4009.28$\pm$0.01 & 4009.21$\pm$0.04 & 4009.24$\pm$0.04 \\
			4026 & 4026.20 & 4026.07$\pm$0.02 & 4026.10$\pm$0.02 & 4026.06$\pm$0.01 & 4026.08$\pm$0.03 \\
			& 4026.36  \\
			4121 & 4120.82 & 4120.93$\pm$0.03 & 4120.86$\pm$0.02 & 4120.89$\pm$0.06 & 4120.89$\pm$0.05 \\
			& 4120.99  \\
			4144 & 4143.76 & 4143.81$\pm$0.01 & 4143.82$\pm$0.01 & 4143.82$\pm$0.02 & 4143.82$\pm$0.02 \\
			4388 & 4387.93 & 4387.93$\pm$0.02 & 4387.96$\pm$0.02 & 4388.01$\pm$0.05 & 4387.97$\pm$0.05 \\
			4471 & 4471.48 & 4471.34$\pm$0.09 & 4471.30$\pm$0.10 & 4471.27$\pm$0.11 & 4471.30$\pm$0.10 \\
			& 4471.68  \\
			4713 & 4713.16 & 4713.20$\pm$0.06 & 4713.38$\pm$0.06 & 4713.19$\pm$0.02 & 4713.26$\pm$0.10 \\
			& 4713.38  \\
			4922 & 4921.93 & 4921.90$\pm$0.18 & 4921.94$\pm$0.15 & 4921.90$\pm$0.12 & 4921.92$\pm$0.15 \\ 
			\hline
		\end{tabular}
	\end{center}
\end{table*}

%________________________________________________________________
\section{Stellar parameters}\label{s_par}

Effective temperatures, \teff, and logarithmic surface gravities, \logg\ (in cm s$^{-2}$),  have been estimated for all our targets. The nature of Be-stellar spectra makes these parameters more difficult to estimate than for normal B-type stars \citep[see, for example,][]{ahm17, dun11}. Indeed at near critical rotational velocities, the conventional model atmosphere assumption of plane parallel (or spherical) geometry breaks down. Then as discussed by, for example, \citet{zor16}, these atmospheric parameters represent some average over the spatially varying photospheric conditions. Additionally the observed spectrum may be contaminated by emission from a circumstellar disc \citep{riv12a, dun11}. Therefore, we have adopted a relatively simple approach, estimating \teff from the spectral type and then the gravity from the higher-order Balmer lines. In turn these have been used to estimate current stellar masses. Further details can be found as supplementary material in Appendix D and in Table \ref{t_atm}.

\begin{table}
	\caption{Estimates of the atmospheric parameters, critical velocities (\vlc) and masses. Effective temperature estimates followed by a colon (:) assume a luminosity class V. The full table is available as supplementary material.}\label{t_atm}
	\begin{center}
		\begin{tabular}{llccccc} 
			\hline\hline			
			VFTS	&	~~\teff	&	\logg	&  \vlc	& M/\Msun  \\
			&           &           &       &          \\
			\hline
			022 & 29600 & 3.60  & 471 & 21   \\
			030 & 19000: & 2.75  & 278 & 18   \\
			034 & 25700: & 4.30  & 585 & 10   \\
			068 & 26200: & 3.40  & 409 & 19   \\
			101 & 27950 & 3.95  & 520 & 14   \\
			\hline
		\end{tabular}
	\end{center}
\end{table}

These stellar parameters can also be used to estimate the stellar critical velocities and again details are provided as supplementary material in Appendix D, where we define three quantities following the nomenclature of \citet{riv12a}. Two of these, \vorb\ and \vc\ are theoretical quantities with the former being the Keplerian circular orbital velocity and the latter this quantity for a star rotating at critical rotation. The third quantity, \vlc, is observational, being deduced from our estimated stellar parameters. Further details are provided in Appendix D and as discussed there:

\begin{equation}
	v_{\rm{orb}}\geq v_{\rm{crit}}\geq v_{\rm{c}}
\end{equation}

Hence \vlc\ estimates (which are listed in Table \ref{t_atm}) will provide lower limits to either the critical, \vc, or the orbital velocities, \vorb.

%________________________________________________________________
\section{Discussion} \label{s_disc}
Below, we compare our estimates of Be-type stellar parameters with those obtained using similar methodologies \citep[][and references therein]{duf12, eva15, gar17}  for other VFTS B-type stars, excluding supergiants and SB2 systems. We also consider equivalent samples of Be-type and B-type stars (again excluding supergiants and SB2 systems) towards the SMC young cluster, NGC\,346. Stellar parameters have been estimated for these targets by \citet{duf19}, using similar methods to those adopted here. However the \vsini-estimates were based on a single feature (either \ion{He}{i} 4026\AA\ or \ion{Si}{iii} 4553\AA) and hence will be less reliable than those for the VFTS samples. Therefore, we have based our discussion principally on the VFTS samples and have mainly used  the NGC\,346 samples to check for consistency.

%________________________________________________________________ 
\subsection{Projected rotational velocities}\label{d_vsini}
Table \ref{t_vsini_mean} summarizes statistics for \vsini-estimates of the Be- and B-type targets in the VFTS and NGC\,346 surveys. The SB1 candidates and those showing no significant radial-velocity variations (henceforth designated as `single-star') are considered separately\footnote{The single-star samples are likely to contain unidentified binaries as discussed by \citet{dun15}}. The Be-type single-star samples have medians (and means) between 240--270\,\kms. Correcting the \vsini-estimates for possible systematic errors as discussed in Sect.\ \ref{s_vsini_bias} would increase these medians by typically 20\,\kms. Similar statistics for the corresponding B-type stars are also summarized in Table \ref{t_vsini_mean} using the estimates in \citet{duf12,duf19}. Their medians and means are lower by typically 100\,\kms\ than the comparable values for the Be-type samples. 

Means for the Be-type SB1 candidates are also lower but are based on small sample sizes. Additionally comparisons of the estimates for single-star and SB1 samples are complicated by the possibility that the latter may be biased to systems with large orbital angles of inclination as discussed by \citet{ram15}. However this would be expected to lead to an overestimation of the mean \vsini-estimate for the binary sample, whilst in fact the opposite is found.
	
Recently several binary systems have been identified with Be-type characteristics and a very narrow lined primary spectrum, implying a projected rotational velocity of $\la$20\,\kms. These include LB1 \citep{liu19, she20, sim20}, HR\,6819 \citep{riv20, bod20, elb21} and NGC2004-115 \citep{len21}. The nature of these systems remains unclear, although it is unlikely that the primary is a typical Be-type star. All the SB1 systems in both of our samples have significantly larger \vsini-estimates (see Table \ref{t_vsini}) and are hence unlikely to be analogs of these systems.

\begin{figure}
	\hspace{-20pt}
	\includegraphics[width=10cm, angle=0]{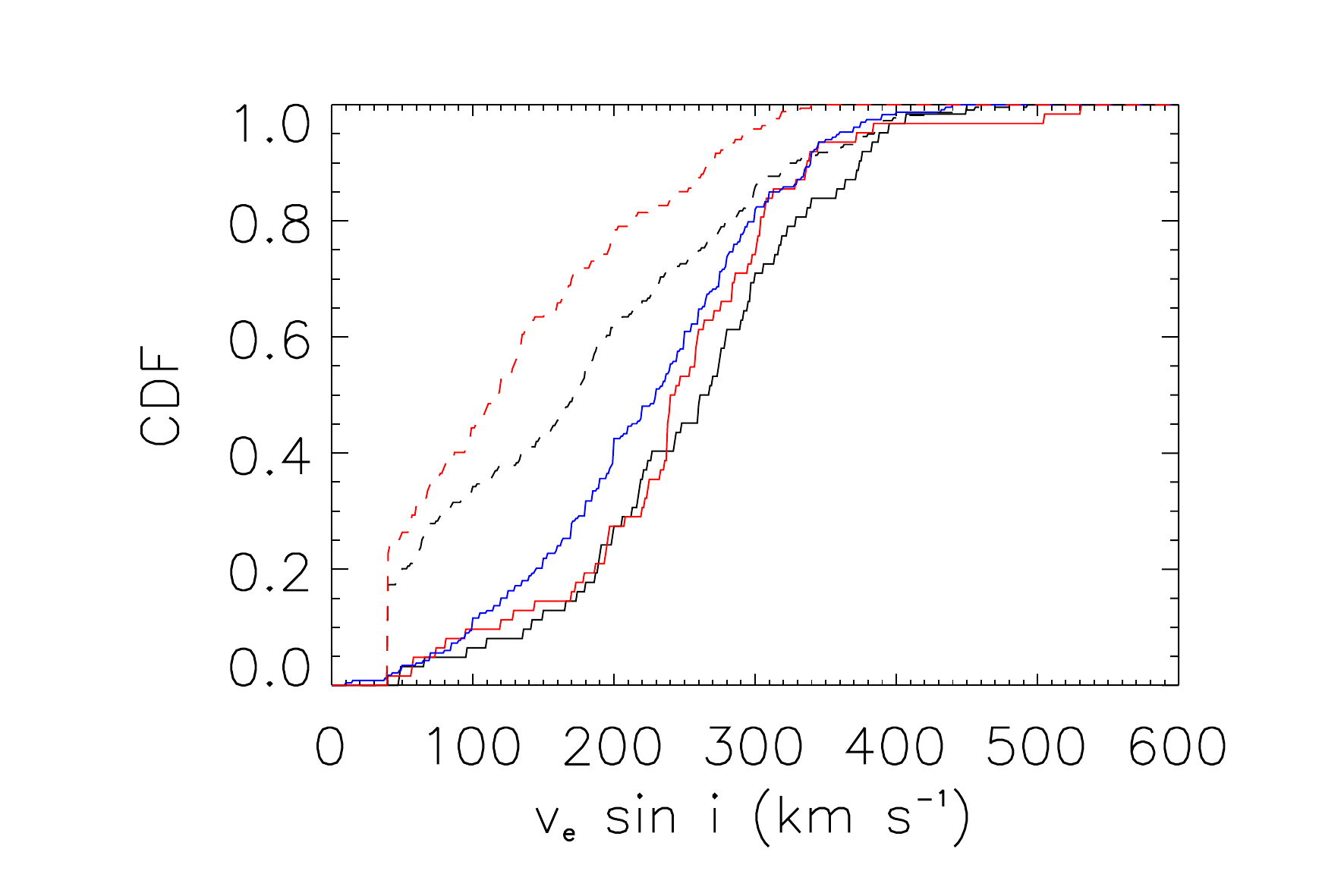} 
	
	\caption{Cumulative distribution function for the VFTS (black lines) and NGC\,346 (red lines) \vsini-estimates. Solid lines are for the apparently single Be-type samples and dashed lines for the corresponding samples of non-supergiant B-type stars. Also shown is the cumulative distribution function for the Galactic Be-type sample of \citet{zor16} (blue solid line).}
	\label{f_cdf}
\end{figure}

In Fig.\ \ref{f_cdf}, cumulative distribution functions (CDFs) are shown for our apparently single B-type and Be-type stellar samples. The apparent difference between the CDFs for our two B-type samples has been discussed by \citet{duf19}. They suggested that it might reflect the larger number of cluster members in the VFTS sample, although the differences were found to be not statistically significant at a 5\% threshold. Also shown in In Fig.\ \ref{f_cdf} is the CDF for the sample of Galactic Be-type stars discussed by \cite{zor16}. 

All the single Be-type single-star samples have a similar distributions, although there is evidence of more targets with lower \vsini-estimates in the Galactic sample. However Kolmogorov-Smirnov (KS) tests \citep{fas87} implied that these differences were not significant at the 10\% level, whilst the selection criteria for the three samples are different. Hence we conclude that there is no significant evidence of rotational velocity differences between the Galactic and Magellanic Cloud Be-type samples.

A KS test returned very small probabilities ($<$10$^{-4}$\%) that our single Be-type and corresponding B-type subgroups are from the same parent population for both the VFTS and the NGC\,346 samples. For the SB1 Be-type systems, KS tests found no statistical significant differences at the 10\% level between either the Be-type stars from the two surveys or between the Be-type samples and their corresponding B-type sample. This is consistent with the small number of SB1 systems in both surveys.

In summary, the means and medians of the \vsini-estimates for Be-type stars in both surveys are larger by typically 100\,\kms higher than those for the corresponding B-type stellar samples. In turn statistical tests imply a very small probability of them having arisen from the same parent population.

\begin{table*}
	\caption{Median and means for the \vsini-estimates for the non-supergiant targets in the VFTS and NGC\,346 surveys. Statistics are given for subgroups with different spectral classifications.}\label{t_vsini_mean}
	\begin{center}
		\begin{tabular}{lllrrrrrrrrrrrccccc} 
			\hline\hline	
			Dataset & \multicolumn{2}{c}{Sample}  & \multicolumn{2}{c}{Measured} 
			& \multicolumn{2}{c}{Fr\'emat} & $n$ \\
			&&                                    & Median & Mean & Median & Mean & $n$  \\
			VFTS    & Be-type & Single            &  268   &  257 & 287    & 282  & 63   \\
            VFTS    & B-type  & Single            &  173   &  173 & 189    & 191  & 219  \\
		    \\
			VFTS    & Be-type & SB1               &  219   &  218 & 237    & 243  & 6    \\
			VFTS    & B-type  & SB1               &  129   &  138 & 145    & 153  & 91   \\
			\\			
			NGC346  & Be-type & Single            &  242   &  245 & 259    & 270  & 62   \\
			NGC346  & B-type  & Single            &  113   &  125 & 129    & 137  & 158  \\  
			\\
			NGC346  & Be-type & SB1               &  229   &  205 & 246    & 222  & 4    \\
			NGC346  & B-type  & SB1               &  137   &  141 & 153    & 155  & 29   \\	
			\hline
		\end{tabular}
	\end{center}
\end{table*}
% Table generated 17/4/20

\subsection{Rotation at near critical velocities}\label{d_near_vc}

Our Be-type samples appear to be rotating at systematically larger velocities than the corresponding B-type samples as has been found in many previous  studies \citep[see, for example,][]{str31, cra05, fre05, dun11, riv12a, zor16}. Indeed it is often assumed that all Be-type stars are rotating at near critical velocities \citep[see, for example,][]{riv12a,bas17}. To test this assumption for our Be-type samples, we have undertaken Monte-Carlo simulations for a Gaussian distribution of rotational velocities: 

\begin{equation}
	\label{e_pg}
	p(\gamma_{\rm{e}})= \frac{1}{\sigma\sqrt{2\pi}}\exp\left( \frac{(\gamma_{\rm{e}}-\gamma_{\rm{cent}})^2}{2\sigma^2}\right)	
\end{equation}

\noindent where $\gamma_{\rm{e}}$\ is the equatorial rotational velocity in units of the critical velocity:

\begin{equation}	
	\gamma_{\rm{e}}= \frac{v_{\rm{e}}}{v_{\rm{crit}}} \label{e_gamma}
\end{equation}

We have adopted $\sigma$=0.02 and three values of $\gamma_{\rm{cent}}$, viz\ 0.7, 0.8, and 0.9 and then simulated the corresponding \vsini-distributions assuming that $\sin i$\ is randomly distributed. Scaling by the VFTS sample size, leads to estimates of the number of targets  (\nc) with projected rotational velocities less than some fraction (\fc) of the critical velocity:

\begin{equation}	
	F_{\rm{crit}}= \frac{v_{\rm{e}} \sin i}{v_{\rm{crit}}}
\end{equation}

\begin{table}
	\caption{Monte Carlo simulations for the number (\nc) of VFTS Be-type stars  that would have a projected rotational velocity (\vsini) less than some fraction (\fc) of the critical velocity (\vc). The simulations assume a Gaussian distribution of equatorial rotation velocities (see Equation \ref{e_pg}) with a $\sigma$=0.02 and centred at \gc\ of 0.7, 0.8 and 0.9.}\label{t_MC}.
	\begin{center}
		\begin{tabular}{cccccc} 
			\hline\hline	
			\gc    & \multicolumn{4}{c}{\fc} \\
			& 0.5 & 0.4 & 0.3 & 0.2 \\
			\hline
			Obs    &  23          & 13           & 7           & 5              \\
			Scaled &  20          & 11           & 6           & 3              \\
			\\
			0.7    & 20.8$\pm$3.8 & 12.4$\pm$3.2 & 6.7$\pm$2.5 & 2.9$\pm$1.7   \\
			0.8    & 15.2$\pm$3.4 & 9.3$\pm$2.8  & 5.1$\pm$2.2 & 2.2$\pm$1.5   \\
			0.9    & 11.6$\pm$3.1 & 7.2$\pm$2.6  & 4.0$\pm$2.0 & 1.7$\pm$1.3   \\
			\hline
		\end{tabular}
	\end{center}
\end{table}
% Table generated 17/1/21

\begin{table}
	\caption{Monte Carlo simulations of the percentage probability of obtaining the  observed number (\nc) of Be-type stars that have a projected rotational velocity (\vsini) less than some fraction, \fc, of the critical velocity, \vc. The simulations assume Gaussian distribution of equatorial rotation velocities with a $\sigma$\ of 0.02 and centred at \gc\ of 0.7, 0.8 and 0.9. Also listed are the centres that best reproduce the \nc\ values (designated 'Implied'). Both sets of \nc\ estimates listed  in Table \ref{t_MC} have been considered.}\label{t_MC_comp}. 
	\begin{center}
		\begin{tabular}{lclllllrr} 
			\hline\hline	
			\vsini\   &  Centre    & \multicolumn{4}{c}{\fc} \\
			scaled & $\gamma_{\rm{cent}}$ & 0.5 & 0.4 & 0.3 & 0.2\\
			\hline
			No  & 0.9    & $<$0.1\% & 2.6\%   & 10.3\%  & 3.1\% \\
			& 0.8    & 2.0\%    & 13.1\%  & 24.5\&  & 7.1\%  \\			
			& 0.7    & 32.4\%   & 47.6\%  & 50.9\%  & 16.8\%  \\
			\multicolumn{2}{l}{Implied $\gamma_{\rm{cent}}$}  & 0.67 & 0.69 & 0.69 & 0.54 \\
			\\
			Yes	& 0.9    & 0.9\%   & 10.5\%  & 20.8\%  & 25.2\%  \\
			    & 0.8    & 10.7\%  & 32.2\%  & 39.6\&  & 38.2\%  \\			
			    & 0.7    & 63.1\%  & 71.8\%  & 67.0\%  & 55.6\%  \\
			\multicolumn{2}{l}{Implied $\gamma_{\rm{cent}}$}  & 0.71 & 0.74 & 0.74 & 0.69 \\			
			\hline
		\end{tabular}
	\end{center}
\end{table}
% Table generated 17/1/21

Table \ref{t_MC} summarizes the predicted means (and standard deviations) of \nc\ for upper limits of \fc\ ranging from 0.5 down to 0.2. Also listed are the observed numbers of VFTS targets using both the measured and the scaled \vsini-estimates listed in Table \ref{t_vsini}. Reasonable agreement with observation is found for \gc$\sim 0.7$. By contrast, all of the observed values are larger than predicted when \gc$\geq$0.8. 

To investigate this further, we have also used our Monte-Carlo simulations to estimate the likelihood that the observed \nc\ values (or larger) would be observed and summarize the results in Table \ref{t_MC_comp}. For both sets of \vsini-estimates,  \gc=0.9 is inconsistent with the observations at a 5\% threshold, whilst adopting \gc=0.8 leads to percentages that are still relatively low. By contrast a value of \gc=0.7 leads to relatively high levels of likelihood, consistent with the predicted and observed \nc\ values being in good agreement.

This comparison implies that not all our targets can be rotating at near critical equatorial rotational velocities. Both \citet{tow04} and \citet{fre05} found that the \vsini-corrections decreased rapidly with decreasing \gae. For example, \citet{fre05} found a typical correction of 10\,\kms\ for a B2\,V star rotating with \gae=0.8. Hence adopting the measured \vsini-estimates for \nc\ may provide a more realistic comparison. Now the simulations with \gc$\geq$0.8 are also inconsistent with the observations at a 5\% likelihood threshold.

An equivalent comparison for the NGC\,346 sample led to similar results. The simulations for \gc$\geq$0.8 are now inconsistent with all the observed \nc-estimates at a 5\% likelihood threshold, with again reasonable agreement being found for \gc$\sim 0.7$

We have used our Monte-Carlo simulations to estimate the \gc\ values which lead to predicted mean numbers of VFTS targets being identical to those observed. These are summarized in Table \ref{t_MC_comp} (designated as 'Implied') and lead to \gc\ values of $\sim$0.7; similar values were found for the NGC\,346 sample.

The simulations have been repeated with $\sigma$\ increased to 0.05. Estimated means and standard deviations of the \nc\ values were both increased by typically 1-2\%, whilst the implied values of \gc\ were effectively unchanged. We have not considered further increases as this would be lead to significant numbers of targets that do not rotate at near critical velocities, inconsistent with the assumption that we were testing.

To investigate the sensitivity of our results to the distribution of rotational velocities, we have also considered a step function, as follows:
	
\begin{equation}
	\label{e_step}
	\begin{split}
	p(\gamma_{\rm{e}}) & = \frac{1}{1-\gamma_{\rm{m}}} \textnormal{~~for~~} 1\geq\gamma_{\rm{e}}\geq \gamma_{\rm{m}} \\
	& = 0 \textnormal{~~otherwise}
	\end{split}
\end{equation}

Monte-Carlo simulations were performed for $\gamma_{\rm{m}}$-values of 0.7, 0.8, 0.9 and the results are summarized in Table \ref{t_MC_step}. The simulations are qualitatively similar to those found adopting a Gaussian distribution, with all of them being inconsistent with the observations at a 5\% statistical level. Additionally the $\gamma_{\rm{m}}$-values required to match the observed values are $\sim$0.5. The implication of these simulations is that any $p(\gamma_{\rm{e}})$-distribution that agreed with the observed \fc-values would have to extend to relatively low \gae-values.

\begin{table}
	\caption{Monte Carlo simulations of the percentage probability of obtaining the  observed number (\nc) of Be-type stars that have a projected rotational velocity (\vsini) less than some fraction, \fc, of the critical velocity, \vc. The simulations assume a step function starting at $\gamma_{\rm{m}}$, with values  of 0.7, 0.8 and 0.9 being considered. Also listed are the  $\gamma_{\rm{m}}$-values that best reproduce the \nc\ values (designated 'Implied'). Both sets of \nc\ estimates listed  in Table \ref{t_MC} have been considered.}\label{t_MC_step}. 
	\begin{center}
		\begin{tabular}{lclllllrr} 
			\hline\hline	
			\vsini\   &  Step position    & \multicolumn{4}{c}{\fc} \\
			scaled & $\gamma_{\rm{m}}$ & 0.5 & 0.4 & 0.3 & 0.2\\
			\hline
		No  & 0.9    & $<$0.1\% & 1.0\%  & 6.4\%  & 2.0\%  \\
			& 0.8    & $<$0.1\% & 2.7\%  & 10.6\&  & 3.1\%  \\			
			& 0.7    & 0.5\%    & 7.4\%  & 18.6\%  & 5.0\%  \\
		\multicolumn{2}{l}{Implied $\gamma_{\rm{m}}$}  & 0.49 & 0.48 & 0.47 & 0.30 \\
		\\
		Yes	& 0.9    & 0.2\%  & 5.5\%    & 14.5\%  & 20.7\%  \\
			& 0.8    & 1.0\%  & 10.7\%   & 21.1\&  & 25.7\%  \\			
			& 0.7    & 4.7\%  & 22.2\%   & 31.1\%  & 32.7\%  \\
		\multicolumn{2}{l}{Implied $\gamma_{\rm{m}}$}  & 0.53 & 0.55 & 0.55 & 0.48 \\			
		\hline
		\end{tabular}
	\end{center}
\end{table}
% Table generated 17/1/21

In summary, the MC simulations are inconsistent with all Be-type stars in our two samples rotating at near critical velocities. Adopting a Gaussian distribution for \gae\ implies  \gc$\sim$0.65--0.75, in good agreement with the mode of the \gae-distribution found by \citet{zor16} for a sample of  Galactic  Be-type stars -- this will be discussed further in Sect.\ \ref{d_gamma}

\subsection{Rotational velocity distributions, \ve}\label{d_ve}

\begin{figure}
	\hspace{-20pt}
	\includegraphics[width=10cm, angle=0]{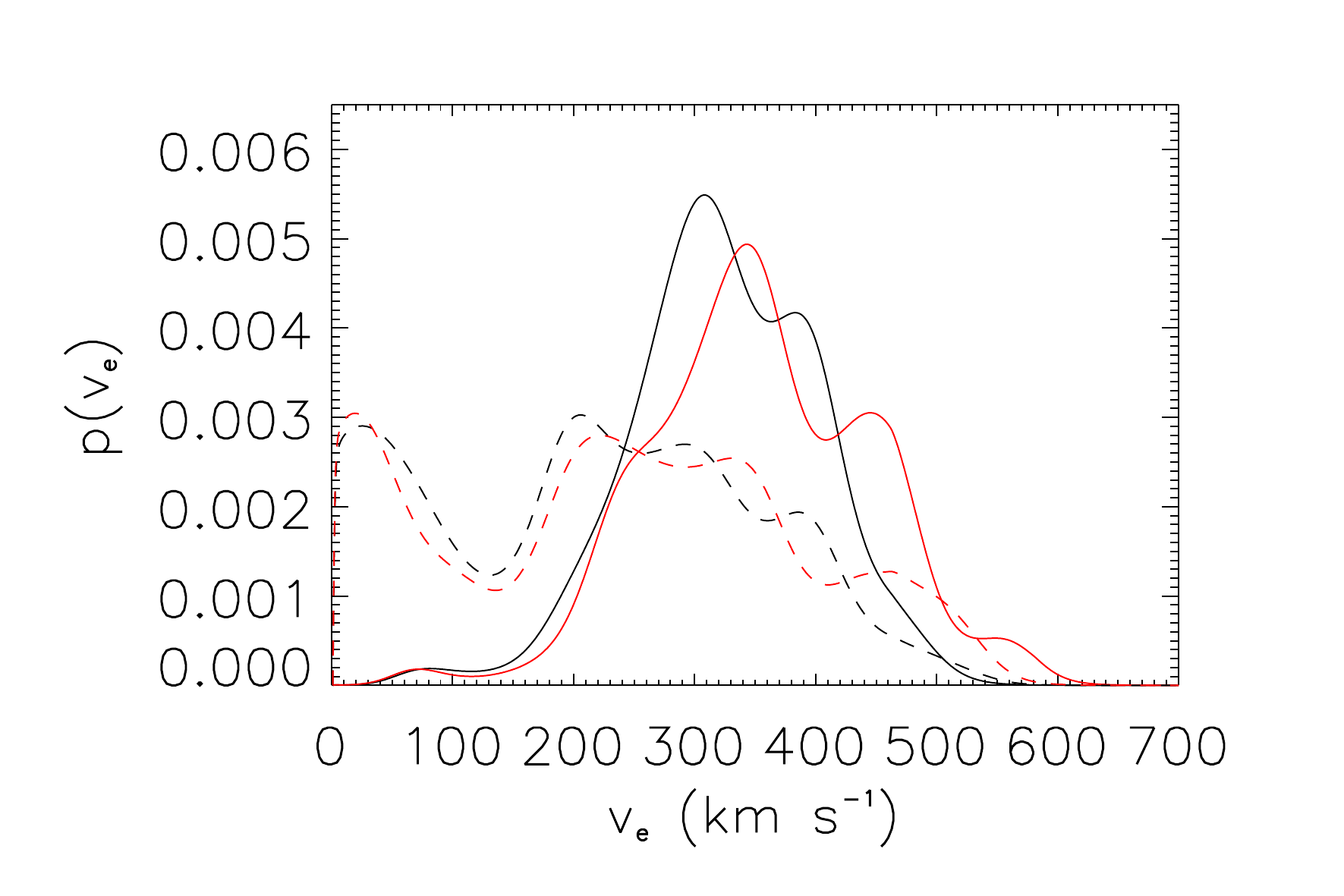}
	
	\caption{Rotational velocity distributions for apparently single VFTS targets. Black and red solid lines are for the Be-type subsample, with the latter including corrections estimated from \citet{fre05}. Dotted lines are analogous results for the B-type sample, excluding Be-type and supergiant targets}
	\label{f_p_ve_VFTS}
\end{figure}

Our sample of single VFTS Be-type contains only 63 targets but can be used to constrain the current distribution of their equatorial rotational velocities. Assuming that their rotation axes are randomly distributed, we can infer the normalised probability density function distribution, $p$(\ve), using the iterative procedure of \citet{luc74} as implemented by \citet{duf12}, where further details can be found. In Fig.\ \ref{f_p_ve_VFTS}, these are shown for both the measured and scaled \vsini-estimates; tests showed that including the 6 SB1 systems did not significantly alter these distributions. Distributions for the single B-type VFTS sample, discussed in Sect.\ \ref{d_vsini} are also shown in Fig.\ \ref{f_p_ve_VFTS}.

Given the relatively small size of the Be-type sample, care should be taken in interpreting these results. In particular, structure with a velocity scale similar to the binning interval of 40 \kms\ used for the observational estimates may not be real. However there does not appear to be a significant number of Be-type stars with equatorial velocities, \ve$\la$150\,\kms. Additionally the maximum equatorial rotational velocity would appear to be in the range $\sim$550-600\,\kms. In Table \ref{t_ve_mean}, we list the median and means for different probability distributions. For the Be-type stars they are in the range 320-350\,\kms, compared with 220-250\,\kms\ for the other B-type stars, consistent with the \vsini-statistics in Table \ref{t_vsini_mean} after scaling by the mean value of $\sin i=\pi$/4 \citep{gra05}. 

\begin{figure}
	\hspace{-20pt}
	\includegraphics[width=10cm, angle=0]{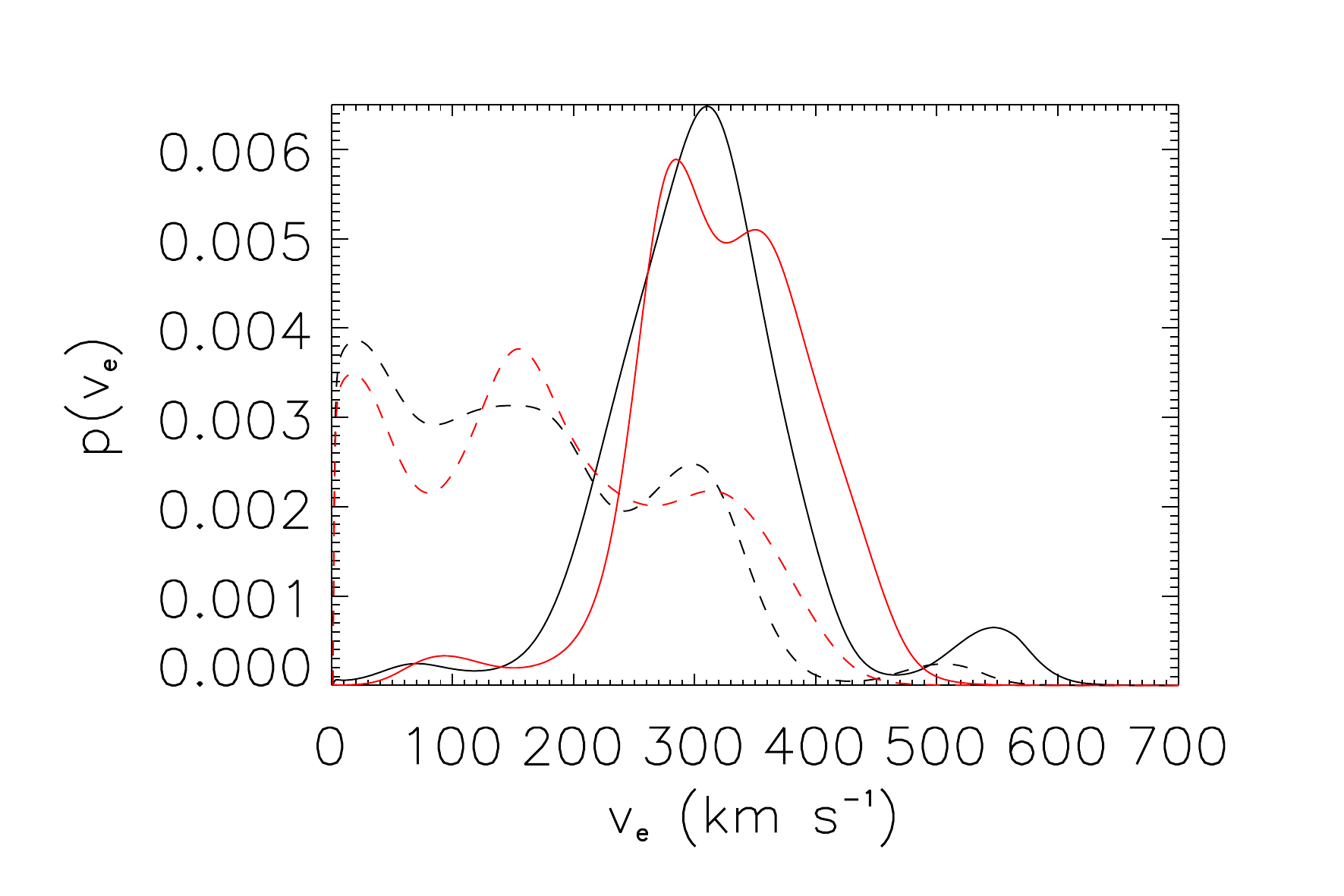}
	
	\caption{Rotational velocity distributions for apparently single NGC\,346 targets. See the caption to Fig.\ \ref{f_p_ve_VFTS} for further details.}
	\label{f_p_ve_346}
\end{figure}

Equatorial velocity distributions have also been inferred using the same methodology for the apparently single B-type and Be-type stars towards NGC\,346 \citep{duf19} and are shown in Fig.\ \ref{f_p_ve_346}. The distribution for the measured \vsini-estimates show a small peak at $\sim$550\,\kms, due to two fast rotating targets (\#1134 and \#1174) that may have unusual evolutionary histories \citep{duf19}. Applying the corrections of \citet{fre05} led to \vsini-estimates ($>$600\,\kms), which were significant larger than the estimated critical velocities; they were therefore excluded from the de-convolution for the scaled estimates. Compared with the VFTS distributions, there again appears to be few (if any) slowly rotating Be-type stars, whilst the upper limit is similar (although possibly slightly lower). The corresponding means and medians for the NGC\,346 samples are also listed in Table \ref{t_ve_mean}. The Be-type values are similar to those for the VFTS samples but those for the B-type sample are lower, as discussed in Sect.\ \ref{d_vsini} for the \vsini-estimates.

In summary, both samples of Be-type stars show an absence of stars with \ve$\la$150\,\kms\ and an upper limit for the equatorial velocity of 500-600\,\kms. They have mean equatorial velocities in the range 310-350\,\kms, whilst there is no convincing evidence for structure in their probability density function distribution, $p$(\ve).

\begin{table}
	\caption{Median and means for the deconvolved rotational velocity distributions for the apparently single non-supergiant targets in the VFTS and NGC\,346 surveys. The columns labebelled 'Fr\'emat' have used the \vsinif-estimates listed in Table \ref{t_vsini}.}\label{t_ve_mean}
	\begin{center}
		\begin{tabular}{lllrrrrrrrrrrrccccc} 
			\hline\hline	
			Dataset & Sample  & \multicolumn{2}{c}{Measured} 
			& \multicolumn{2}{c}{Fr\'emat}  \\
			&                            & Median & Mean & Median & Mean  \\
			VFTS    & Be-type            &  323   &  325 & 349    & 353   \\
			
			VFTS    & B-type             &  226   &  223 & 246    & 243   \\
			\\	
			NGC346  & Be-type            &  305   &  309 & 323    & 323   \\
			
			NGC346  & B-type             &  155   &  168 & 169    & 181   \\  
			\hline
		\end{tabular}
	\end{center}
\end{table}

\subsection{\gae-distributions, $p$(\gae)} \label{d_gamma}

\begin{figure}
	\includegraphics[width=9cm, angle=0]{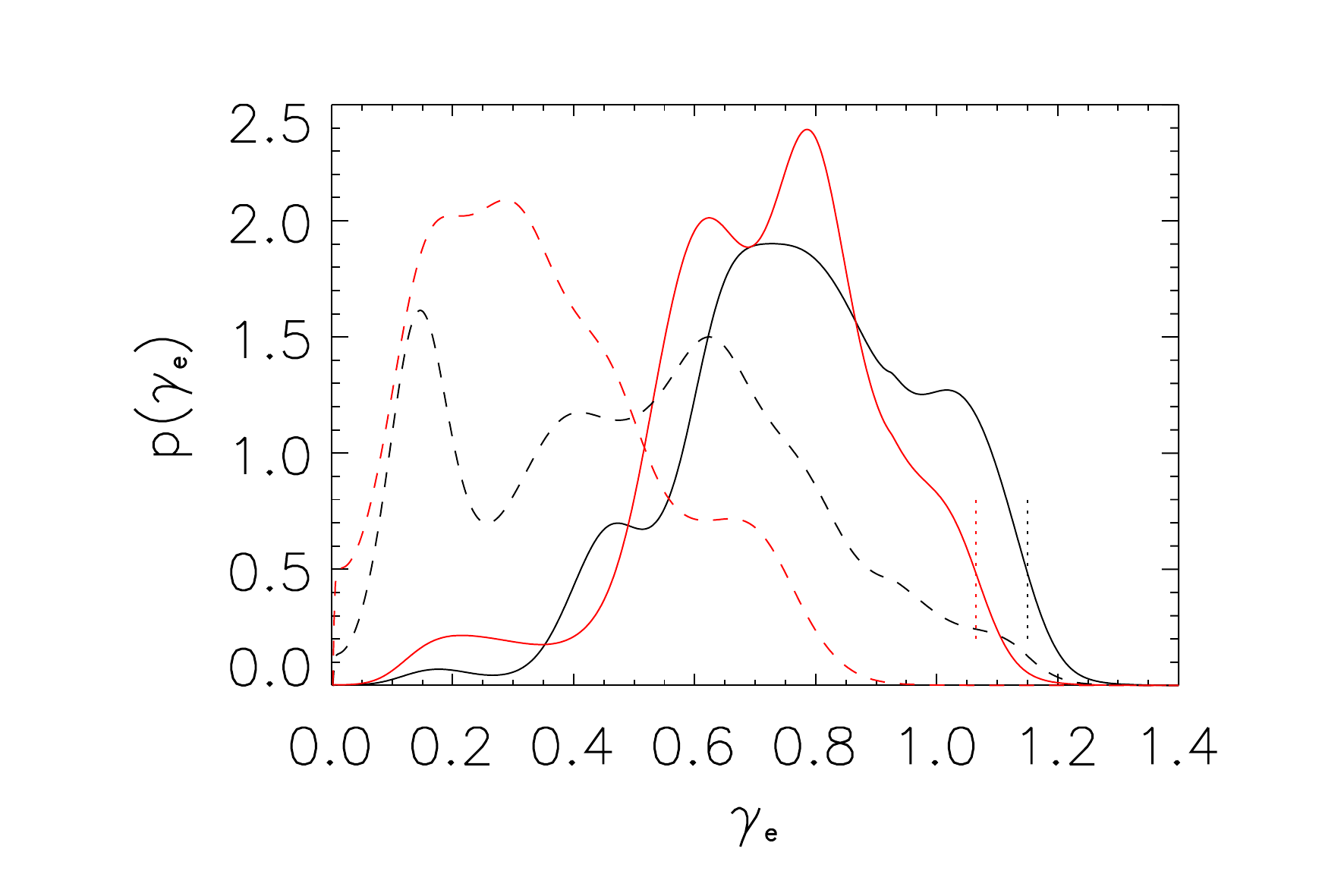}
	
	\caption{\gae-distributions for apparently single Be-type targets. Black and red solid lines are for the the VFTS and NGC\,346 samples respectively. Dotted lines show the adopted maximum values for the distributions as discussed in Sect.\ \ref{d_gamma}. Dashed lines are the distributions for the analagous B-type samples.}
	\label{f_p_gamma}
\end{figure}

The normalised probability distribution, $p$(\gae), introduced in equation \ref{e_pg} can be inferred by deconvolving the \fc\ values deduced from the \vsini\ and \vlc\ estimates. The uncorrected estimates, listed in Tables \ref{t_vsini} and \ref{t_atm} respectively, were initially adopted but we discuss this further below. The de-convolved probability distributions for the VFTS apparently single Be-type dataset is shown in Fig.\ \ref{f_p_gamma}, together with that deduced for the NGC\,346 sample. We have also performed similar de-convolutions for the corresponding B-type samples introduced in Sect.\ \ref{d_vsini} and these are also shown in Fig. \ref{f_p_gamma}. The \vlc-estimates utilised the stellar parameters in \citet{sch18} for the VFTS sample and in \citet{duf19} for the NGC\,346 sample.

The two probability distributions for the Be-type samples have similar characteristics with a dearth of targets at \gae$\la$0.4 and extending to \gae$>$1, which would imply equatorial rotational velocities in excess of the critical velocity.  For the VFTS, this would constitute $\sim$17\% of the Be-type sample, increasing to $\sim$27\% after applying the \vsini-corrections of \citet{fre05}. The percentages for the NGC\,346 sample would be 7\% and 14\% respectively. By contrast the B-type samples have a significant number of targets at small \gae-values, with $<4$\% of the VFTS sample and $<1$\% of the NGC\,346 having \gae$>$1.

\citet{zor16} undertook a similar de-convolution for Galactic Be-type stars and found approximately 4\% of their sample also had \gae$>$1. They considered the systematic biases on their \vsini\ and \vc\ estimates on a star-to-star basis and generally found that their \fc\ estimates were reduced. These biases are the most likely explanation of the extension of our \gae-distributions beyond \gae>1.

We have attempted to estimate the overall effect of these biases by identifying upper limits, \gmax\ for the \gae-distibutions, We have adopted a limit at which 98\% of the \gae-distributions would have \gae$<$\gmax. This leads to \gmax\ estimates of 1.150 and 1.064 for the VFTS and NGC\,346 samples respectively and these are illustrated in Fig.\ \ref{f_p_gamma}. Scaling our distributions by these estimates would lead to median (and means) for both our datasets of $\sim$0.68 in excellent agreement with the estimate of \citet{zor16} of 0.64-0.68 for their Galactic sample. Our choice of \gmax\ is of course subjective but clearly some adjustment is required to ensure a physically realistic distribution. For the VFTS and NGC\,346 B-type samples, the corresponding \gae-medians would be $\sim 0.51$\ and $\sim 0.32$\ respectively, reflecting the lower estimates of the \ve-means of the latter shown in Table \ref{t_ve_mean}.

Recently \citet{bal20, bal21} have discussed TESS photometry of over 400 Galactic Be-type stars. The majority are short term variables and their \vsini-estimates imply that their fundamental periods reflect rotational modulation. In turn these allow estimates of their equatorial velocities and hence of $p$(\gae) \citep[see Fig.\ 5 of ][]{bal21}, which is similar to that of \citet{zor16} and to those shown in Fig.\ \ref{f_p_gamma}. For example there is a lack of Be-type stars with \gae$\la$0.3 and then a broad distribution up to \gae$\sim 1$. For early Be-type stars, their mean \gae-value is $\sim$0.65 in excellent agreement with the values discussed above. However, although the methodology of \citet{bal20, bal21} is independent of those used by \citet{zor16} and here, it depends on an identified frequency  being due to rotational modulation. As discussed by \citet{lab20} and \citet{bar22}, the frequency spectra of the TESS photometry for Be-type stars are complex and diverse with essentially all targets being variable over different time cadences. Additionally they identify several other mechanisms that may lead to variability besides rotational modulation.

In summary, the deconvolved \gae-distributions for both Magellanic Cloud datasets are similar and imply a dearth of targets with \gae$\la$0.4, with a broad distribution reaching up to critical rotation, \gae$\simeq$1. Our best estimate for the mean or median of the \gae-distribution is $\sim$0.68. These characteristics are in agreement with those found by \citet{zor16} and \citet{bal20,bal21} for Galactic samples. 

\subsection{Rapidly rotating Be- and B-type stars} \label{d_fast}

The \ve-distributions in Figs.\ \ref{f_p_ve_VFTS} and \ref{f_p_ve_346} imply that there are a significant number of apparently single rapidly-rotating B-type stars that do not currently exhibit Be-type characteristics. We have estimated the relative numbers of rapidly rotating Be-type and B-type stars in our two samples using lower limits to the equatorial rotational velocity of 200\,\kms and 300\,\kms. Essentially all ($\sim$95\%) Be-type stars were included using the former limit and between 50-70\% for the later. The percentages were lower for the B-type stars ranging from $\sim$20-60\%. To estimate the actual number of targets, these percentages must be scaled by the number of targets and these estimates are summarized in Table \ref{t_high_ve}. For both datasets, there are a significant number of rapidly rotating B-type stars, leading to a Be-type to B-type ratio ($R$) of $\sim$0.5--0.7 for the VFTS sample and $R\sim$1.0--1.5 for the NGC\,346 sample. Apart from stochastic uncertainties, this variation in $R$\ could reflect intrinsic differences between the two samples that are discussed as supplementary material in Appendix E.

Using the inferred \gae-distributions shown in Fig.\ \ref{f_p_gamma}, we can undertake a similar comparison in terms of \gae. We have considered lower limits for \gae\ of 0.6 and 0.8 and summarize the results in Table \ref{t_high_ve}. For the VFTS sample, the number of B-type stars is similar to or greater than that for the Be-type stars, in agreement with the the comparison for the \ve-distribution. By contrast, the Be-type stars are more numerous in the NGC\,346, especially for the case with \gae$\geq$0.8. However it should be noted that this sample relates to only the $\sim$30\% of the Be-type stars with the largest  \gae-values. Additionally we have not corrected the \gae-distributions to account for the significant number of Be-type stars with \gae$>$1.

For the VFTS sample (which is better constrained in terms of target selection, environment and binarity identification), the B-type stars are more prevalent than the Be-type stars, when considering either the \ve-distributions or the \gae-distributions. This is consistent with the magnitude limited samples of Be-type and Bn-type stars drawn from the Bright Star Catalogue \citep{bsc} and discussed by \citet{riv12a}. Given the transitory nature of the Be-type phenomena \citep[see, for example,][]{riv12a}), some of the rapidly rotating B-type stars may be quiescent Be-type. Unfortunately, statistics for the duration of the Be-type and quiescent phases are not available and hence it is not possible to estimate the number of rapidly rotating B-type stars that are quiescent Be-type.

\begin{table*}
	\caption{Estimated numbers of Be-type and apparently single B-type stars with equatorial rotational velocities (\ve) greater than 200 or 300\,\kms. Results are obtained using the probability distributions, p(\ve) discussed in Sect.\ \ref{d_ve}, scaled by the sample size, $N$. Comparable estimates are also provided for the number of targets with \gae-values greater than 0.6 and 0.8, using the probability distributions, p(\gae) discussed in Sect.\ \ref{d_gamma}.}\label{t_high_ve}
	\begin{center}
		\begin{tabular}{lllrrrrrrrrrrrccccc} 
			\hline\hline	
			Dataset & Sample  & $N$ & \multicolumn{2}{c}{Measured} 
			& \multicolumn{2}{c}{Fr\'emat} & \multicolumn{2}{c}{\gae}  \\
			&&                            & $>$200 & $>$300 & $>$200 & $>$300 &  $>$0.6 & $>$0.8 \\
			VFTS    & Be-type &  69   &  65  & 43   & 64    & 48 & 57 & 32   \\	
			VFTS    & B-type  & 219   &  126 & 66   & 127   & 80 & 86 & 32   \\
			\\	
			NGC346  & Be-type &  66   &  62  & 35   & 63    & 40 & 50 & 23   \\
			NGC346  & B-type  & 158   &  58  & 23   & 64    & 30 & 21 & 2    \\  
			\hline
		\end{tabular}
	\end{center}
\end{table*}

\subsection{Binarity} \label{d_bin}

Relatively few binary Be-type candidates have been identified in either the VFTS or NGC\,346 datasets. This is consistent with a recent study of multiplicity in Galactic \citep{bod20} and SMC \citep{bod21} Be-type samples. The latter utilised MUSE spectroscopy of the core of NGC\,330 and found that the observed spectroscopic binary fraction of Be-type stars (2$\pm$2\%) was lower than that for B-type stars (9$\pm$2\%).

The time cadence of our NGC\,346 observations was limited with between two and four epochs covering a period of 6 weeks or less. Additionally the adopted methodology \citep{duf19} would only have identified systems with relatively large radial velocity variations. Hence, here, we concentrate on the VFTS targets that typically had 6 epochs with the time cadence \citep{eva11} such that the predicted detection probability for massive primaries having secondaries with a mass ratio, $q>0.1$, was greater than 60\% for periods of $\leq$100\,d \citep{san13, dun15}. 

The accuracy of the radial velocity estimates depended on, inter alia, the S/N of the spectroscopy and the degree of rotational broadening. For the VFTS sample, the standard deviations for the radial velocity estimates for a given target and epoch covered a wide range from $<1$\kms to $\sim$50\kms\ with a median of 6.8\,\kms. The primary criterion for binarity was that variations in the radial velocity estimates were significant at a 4$\sigma$-level. Excluding supergiants, all identified B-type binaries had a range of radial velocity estimates, $\ga$20\kms. For the VFTS Be-type sample, the median of the standard deviations for the radial velocity estimates for a given target and epoch is 11.0\,\kms. The increase over that for the B-type sample reflects the greater rotational broadening in the Be-type sample.

The lower fraction of {\em detected} binary in the VFTS Be-type sample compared with that for the B-type sample could be due to real differences. Alternatively it could just reflect the intrinsic difficult of estimating radial velocities for Be-type stars, together with the relatively small sample size. To test the latter, we list in Table \ref{t_bin}, the number of VFTS Be-type and B-type SB1 candidates using the different sets of lines summarized in Table \ref{t_lines}. The Be-type statistics exclude the target \#135 for the reasons discussed in Sect.\ \ref{s_bin}; those for the B-type targets were taken from \citet{dun15} and assigned to the different line-sets using \citet{duf12}.

\begin{table}
	\caption{Number of targets (n) and binary candidates (SB1) in the VFTS Be-type and B-type samples. Also listed are the p-statistic from $\chi^2$\ and Fischer statistical tests.}\label{t_bin}
	\begin{center}
		\begin{tabular}{llllll} 
			\hline\hline	
			Line-set & Sample   & n    &  SB1 & p($\chi^2$) & p(Fischer) \\
			1        & Be-type        & 11   &  2  \\			
			& B-type         & 137  & 51   & 0.20        & 0.33   \\
			\\
			2        & Be-type        & 58   &  4  \\			
			& B-type         & 173  & 40   & 0.0065      & 0.0062    \\
			\\
			Both     & Be-type        & 69   &  6  \\			
			& B-type         & 310  & 91   & 0.0037      & 0.0028   \\
			\hline               
		\end{tabular}
	\end{center}
\end{table}

Only 11 Be-type stars were analysed using line-set 1 of which two are SB1 candidates, equating to $\sim$20\%; for the 137 B-type targets, 51 stars were identified as SB1 candidates, corresponding to $\sim$40\%. However as can be seen from Table \ref{t_bin} high probabilities are returned that these are from the same population using both a $\chi^2$\ or Fischer test \citep{mod11}. For line-set 2, the sample sizes are larger and lead to observed binary fractions of $\sim$7\% and 23\% for the Be-type and B-type samples respectively. Now the statistical tests imply that the difference in the observed binary fraction is significant. For completeness, we have also considered the complete samples (both line-sets 1 and 2) and again the statistical tests imply that the binary fraction in the two samples differ.

The above analysis does not take into account the larger uncertainties in the Be-type radial velocity estimates. We have attempted to investigate this as follows. As discussed above, the threshold for identifying binaries was approximately three times the median uncertainty in the radial velocity estimates. Excluding B-type binaries with ranges in radial velocity estimates less than three times the median {\em Be-type} uncertainty would decrease the number of B-type binaries identified to 71. In turn this would lead to p-statistics of 0.036 ($\chi^2$) and 0.040 (Fischer). These are still both statistically significant at a 5\% level implying that the different binary fractions are not due to different observational uncertainties. We emphasize that this comparison is only relevant to differences in the binary populations for systems that would have been detectable for the VFTS sample. As discussed above this would be for massive binaries with periods, $P\la$100\,d and with mass ratios, $q\ga$0.1.

If the binary fraction of such Be-type stars was smaller than that of the normal B-type stars, then one consequence could be a smaller measured radial velocity dispersion for the Be-type stars.  While the VFTS data are well suited to this task, the full B-type stellar sample in the 30 Dor region has a large velocity dispersion, $\sim$20\,\kms, and significant velocity structure, that could obscure any potential difference between Be-type and non-Be-type stars.  However the VFTS survey region contains two older clusters, Hodge 301 and SL639, both of which have main sequence early B-type stars with well defined cluster velocities. As the sample sizes for each cluster are small, we combine them after shifting their stellar radial velocities to the rest frame cluster velocities \citep{pat19}, giving a total of 11 B-type stars, and 16 Be stars. 

\begin{figure}
	\includegraphics[width=9cm, angle=0]{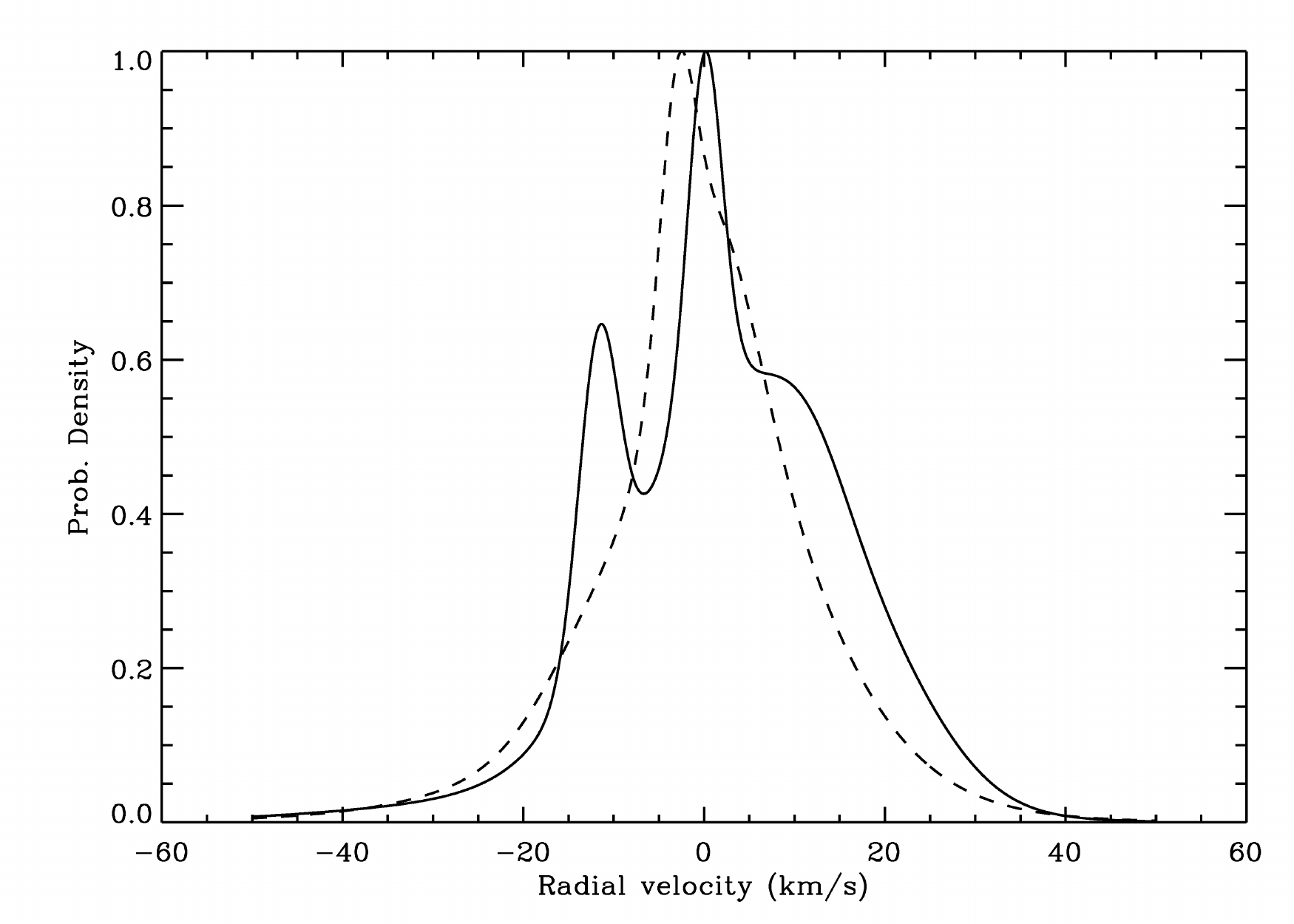}	
	\caption{Combined probability densities for the radial velocity estimates of the Be-type (dotted line) and B-type (solid line) targets in Hodge 301 and SL639.}
	\label{f_v_r}
\end{figure}

Normalized probability density plots are shown in Fig.\ \ref{f_v_r} and hint at a smaller velocity dispersion for the Be-type stars, with weighted sample standard deviations of 8.3\,\kms for the B-type stars, but only 3.5\,\kms for the Be stars. The methodology for the B-type radial velocity estimates differed from that for the Be-type stars as it used Gaussians in the profile fitting and did not include systematic offsets in the \ion{He}{i} lines discussed in Appendix C. For the former, targets with low \vsini-estimates that utilised line-set 1 yielded central wavelength that agreed to within 0.01\AA\ irrespective of the adopted profile. Targets utilising line-set 2 showed differences of typically 0.05\AA, equating to 3--4\,\kms. However the variations were not systematic implying that the uncertainties in the mean radial velocity estimates would be smaller.  The choice of either the laboratory wavelengths or those adopted here (see Table \ref{t_vr_wave}) typically changed the mean radial velocity estimates by 3--4\,\kms but as these changes were systematic the changes in the velocity dispersion would be smaller. 

Given these uncertainties and the small sample sizes, the differences in the velocity dispersion from the two samples may not be significant. However they imply that there is a smaller SB1 Be-type fraction with orbital parameters similar to those of the B-type SB1 population. This would be consistent with either Be-type stars being formed by a single star evolutionary channel or being the products of post-transfer (case-B) binaries. The latter has been discussed by \citet{sha14, lan20}, who predict periods of the order of a few 100 days to a few years. Such binaries would have be difficult to detect with the VFTS time cadence \citep{dun15}, especially given the relatively large projected rotational velocities.

\begin{figure}
	\includegraphics[width=9cm, angle=0]{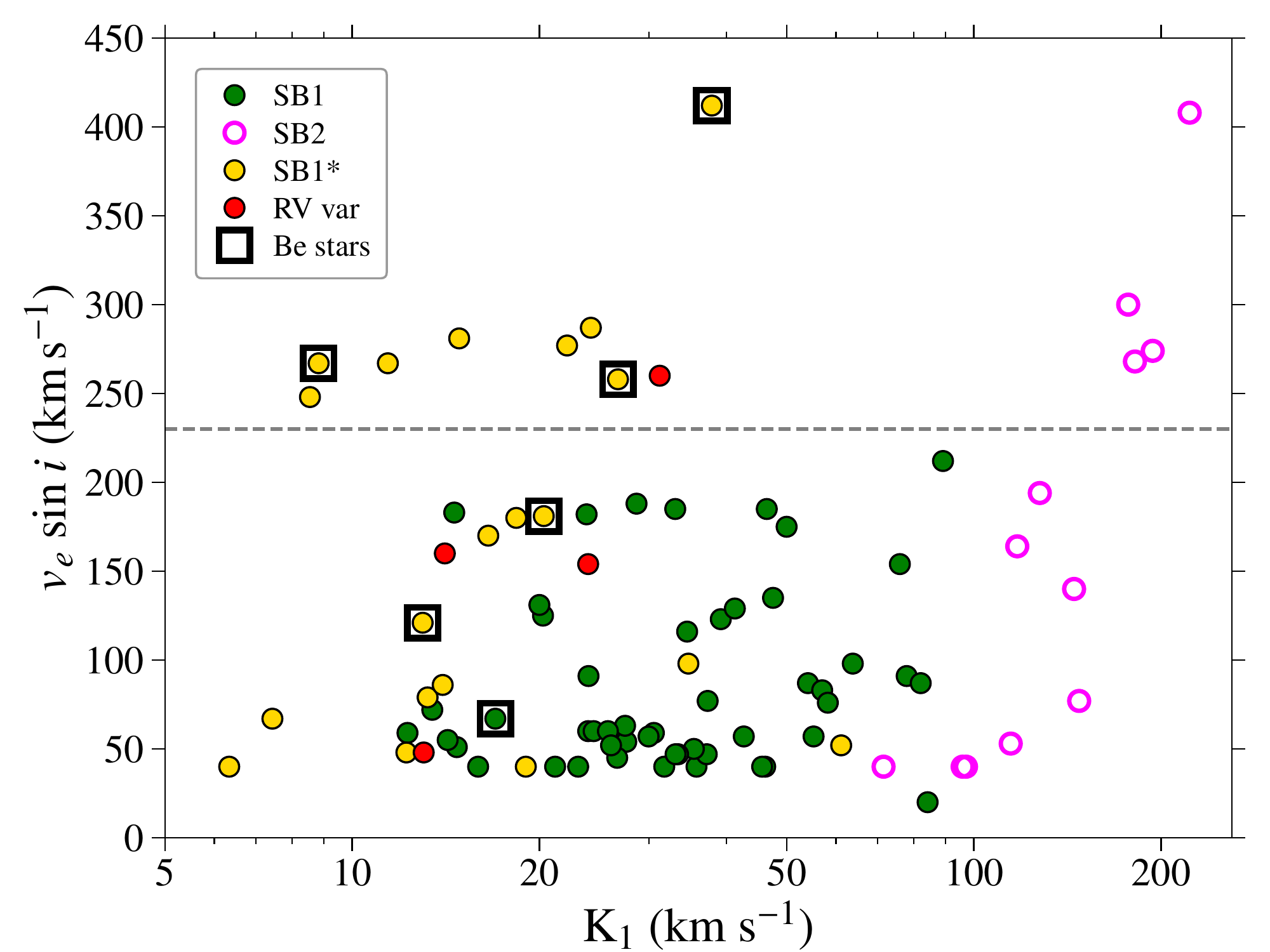}	
	\caption{Estimated semi-amplitudes for VFTS binary candidates taken from \citet{vil21} plotted against their estimated projected rotational velocities. Be-type stars are shown with black boxes.}
	\label{f_K1}
\end{figure}

Recently multi-epoch spectroscopy has been obtained for 88 VFTS B-type binary candidates \citep{vil21}. Binarity was confirmed in 64 targets corresponding to 73\% of the sample with 50 SB1 and 14 SB2 systems. For another 20 targets, designated as SB1$^*$, clear signs of periodicity were identified but it was not possible to definitively confirm binarity. The remaining 4 targets showed evidence of radial velocity variations but not of periodicity and were designated as `RV var'. In Figure \ref{f_K1}, the estimated semi-amplitudes ($K_{\rm{1}}$) are plotted against the (uncorrected) \vsini-estimates. 

For our 6 SB1 Be-type candidates, one (\#874) was confirmed as an SB1 system, with the others  being designated as SB1$^*$\ (see Sect.\ \ref{s_bin} for details); their orbital parameters are summarized in Table \ref{t_sb1}.  They have estimated periods from $\sim$1 day to $\sim$1 year and eccentricities between 0.20 and 0.64 whilst their systemic radial velocities are consistent with the other B-type systems. The periods are smaller than have been found in Galactic B-type stars \citep{oud10, riv12a} and may in part reflect the detection sensitivity for the VFTS sample, decreasing for periods, P$\ga$100 days. However the shortest period binaries may not be able to contain a Be-type disc within their Roche lobe \citep{riv12a, len21}, implying that they may not be classical Be-type stars. It is also notable that all the Be-type systems have large eccentricities.
	
	\begin{table}
		\caption{Parameters for the VFTS Be-type systems with orbital parameters taken from \citet{vil21}}\label{t_sb1}
		\begin{center}
			\begin{tabular}{clllllllllll} 
				\hline\hline	
				VFTS&   ~~P     &   ~~K$_1$ &   ~~$e$   &   ~~$f(m)$        \\
				&   days    &   \kms    &           &   \Msun               \\    
				213 &   13.6    &   20.3    &   0.64    &   0.00536         \\
				337 &   25.5    &   37.9    &   0.55    &   0.08435         \\
				697 &   1.49    &   12.8    &   0.31    &   0.00029         \\
				847 &   1.23    &   26.7    &   0.50    &   0.00159         \\
				874 &   370.8   &   17.0    &   0.21    &   0.17616         \\
				877 &   94.7    &   8.8     &   0.32    &   0.00572         \\
				\hline               
			\end{tabular}
		\end{center}
	\end{table}
	
Inspection of Fig.\ \ref{f_K1} also shows that their estimated semi-amplitudes are all relatively low, with $K_{\rm{1}}<40$\,\kms; by contrast \citet{vil21} found 29 VFTS B-type systems (that have not been classified as supergiants), which had $K_{\rm{1}}>40$\,\kms. The number of VFTS targets classified \citep{eva15} as Be-type and B-type (excluding supergiants) were 69 and 310 respectively. Fischer and $\chi^2$\ tests returns a probability of $<$1\% that the Be-type and B-type samples have the same frequency of binary systems with $K_{\rm{1}}>40$\,\kms.  Hence we conclude that there is some evidence for a dearth of spectroscopic VFTS Be-type binaries, whilst those that are observed have significantly lower semi-amplitudes than those for the B-type binaries.

\section{Principal results and implications}\label{d_con}

The main results of our analysis are as follows:

\begin{enumerate}
	\item For both of our samples, the \vsini-estimates of the Be-type are larger than those for normal B-type stars. The means for the former are typically 100\,\kms\ larger than those for the latter.
	\item Monte-Carlo simulations imply that for our Be-type samples, the mean ratio of the rotational velocity to the critical velocity is \gae$\sim$0.7.
	\item The de-convolution of the \vsini-estimates leads to mean \ve-estimates of 320--350\,\kms and 305--325\,\kms\ for the VFTS and NGC\,346 samples respectively. These are 100--150\,\kms\ larger than those for the corresponding B-type samples.
	\item The deconvolution of the \fc\ estimates lead to \gae-distributions that have a dearth of targets with \gae$\la$0.4, with a broad distribution reaching up to critical rotation, \gae$\simeq$1. Our best estimate for the mean or median of the \gae-values is 0.68, consistent with the Monte-Carlo simulations. There is no evidence for any significant differences between the SMC and LMC samples, whilst these characteristics are very similar to those found by \citet{zor16} and \citet{bal21} for Galactic Be-type stars.
	\item Rapidly-rotating B-type stars appear to be as numerous as their Be-type counterparts. The ratio of the number of Be-type stars to B-type stars, $R$ is estimated as $\sim$0.5--0.7 and $\sim$1.0--1.5 for the VFTS and NGC\,346 samples respectively.
	\item The frequency of Be-type stars identified as binary systems  is lower than that for normal B-type stars. For the VFTS samples the semi-amplitudes for the former are also lower than those of the latter with a high level of significance, which is consistent with the differences in their radial velocity dispersions.
\end{enumerate}

Below we discuss our results in the context of the current evolutionary models for the formation of Be-type stars. These fall into two main categories, viz.\ that Be-type star have evolved either as single stars or as multiple system with mass transfer (henceforth termed `single' and `binary'). The former, first suggested by \citet{str31}, has been reviewed by \citet{riv12a} with extensive modelling and population synthesis recently undertaken by \citet{has20}. The latter has been discussed by \citet{pol91, deM13, sha14, sha20, lan20} and for example, is clearly relevant to Be/X-ray binaries \citep[see, for example,][]{rag05}. Recently it has been invoked to explain the candidate intermediate mass black hole systems, LB1 and HR6819 \citep{she20, bod20a, elb21} as Be-type stars plus a stripped helium star (henceforth characterised as Be/He binaries). 

The wide range of \gae-values found in Sect. \ref{d_gamma} and illustrated in Fig.\ \ref{f_p_gamma} provide a challenge for both evolutionary histories. For single star evolution, \citet{has20} find that reasonable agreement for the fraction of Be-type stars in the B-type population can be obtained if Be-type stars have \gae-values in the range 0.7--0.8. The \gae-distributions illustrated in Fig.\ \ref{f_p_gamma} imply that approximately 60\% of Be-type stars have \gae$\geq$0.7; scaling these distributions as discussed in Sect.\ \ref{d_gamma} would reduce this percentage to $\sim$48\%. However \citet{has20} assumed that all B-type stars with a \gae-value above a given threshold would exhibit Be-type phenomena and as discussed in Sect.\ \ref{d_fast}, this is not the case. In order to produce sufficient Be-type stars would require a lowering of the \gae-threshold in their population synthesis. In turn this would lead to better agreement with the \gae-distributions shown in Fig.\ \ref{f_p_gamma}. Hence given the uncertainties in both our results and the simulations of \citet{has20}, our \gae-distributions may be consistent with a single star evolutionary history.

Many authors \citep[see, for example,][]{pol91, deM13, sha14, sha20, lan20} have discussed the evolution of  massive-star binaries until the primary becomes either a white dwarf, a stripped helium star, a neutron star or a black hole. These simulations imply that the secondary's rotational velocity will increase due to the accretion of mass and angular momentum from the primary. There will also be mechanisms that `spin-down' the secondary, with \citet{sha20} finding that tidal interactions are especially important. \citet{lan20} considered the evolution of binary system leading to a black hole and a massive main sequence star. For the latter, they deduced the probability distribution, p(\gae), which  monotonically increased from \gae$\sim$0.2 up to unity; Be-type stars with neutron star or white dwarf companions would probably suffer less spin down (N.\ Langer, private communication). As such, their \gae-distribution would be shifted to higher values of \gae, consistent with the observed Be-type stellar distribution shown in Fig.\ \ref{f_p_gamma}. 

Very recently, \citet{has21} have investigated mass transfer interactions in B-type stars that could lead to Be-type secondaries. They focused on Be-type stars in clusters and deduced an expression for the fraction of Be-type to the total initial binary population in terms of the stellar mass compared with the turn-off mass of the cluster. They concluded that `binary evolution does not allow more than around 30\% of stars to have been spun up through binary interaction and become emission line objects'. For the VFTS sample, 73 Be-type stars have been identified out of a B-type population of 438 targets \citep{eva15}. Excluding the 52 B-type supergiants (which will normally have evolved from O-type progenitors) leads to 386 targets. Making the extreme assumption that the initial binary fraction, $f$, was unity would then lead to Be-type constituting 19\% of the binary population. \citet{dun15} estimated that the current binary fraction (allowing for unidentified systems) was $f=0.55\pm0.11$ and this would increase the Be-type population to 28-43\% of the binaries. These latter percentages should be treated with caution because, as discussed by \citet{has21}, the initial binary fraction and that currently observed may differ significantly. For NGC\,346, the Be-type fraction of the total B-type population (again excluding supergiants) would be 23\% with the current binary fraction being poorly constrained. Hence the frequency of Be-type stars within our samples are consistent with the simulations of \citet{has21}. However, \citet{bod20b} find a {\em lower limit} on the Be-type stellar fraction of $\sim32$\% in the older SMC cluster NGC\,330, which is larger than the predictions of \citet{has20}

\citet{bod20} investigated a sample of 287 Galactic early-Be-type targets and concluded that there were 'no confirmed reports of Be binaries with MS companions' and that the Galactic sample `strongly supports the hypothesis that early-type Be stars are binary interaction products that spun up after mass and angular momentum transfer from a companion star'. However their analysis was complicated by the possibility of significant biases, which they characterised as the 'vast heterogeneity of techniques, data quality, and focus' of their sample. Recently \citet{bod21} have investigated binarity for B-type stellar population in the SMC cluster NGC\,330 and again found a smaller observed fraction amongst the Be-type stars. These analyses are consistent with the dearth of spectroscopic binaries found in the VFTS Be-type sample discussed in Sect.\ \ref{d_bin}.

From the above discussion, it is not possible to conclude definitively that our Be-type samples are produced predominantly through one evolutionary pathway. The wide range of \gae-values found in both of our samples (and indeed in other studies) requires a mechanism that leads to targets which have {\em currently} relatively low \gae-values. If Be-type stars were the products of binary evolution, the substantial populations of rapid rotating B-type stars could then be due to the latter being single stars. Additionally recent modelling of binary populations is consistent with the observed number of Be-type stars in our sample. Assuming that one mechanism did dominate the production of classical Be-type stars, our results would favour a binary evolutionary history. 

\section*{Acknowledgements}

We are grateful to Michael Abdul-Masih, Nate Bastian, Julia Bodensteiner, Ben Hastings, Norbert Langer, Fabian Schneider and Jorick Vink for useful discussions. The paper has also benefited from constructive comments from the referee. Based on observations at the European Southern Observatory Very Large Telescope in programmes 171.D-0237171.0237, 182.D-0222 and 096.D-0825 and on observations made with the NASA/ESA Hubble Space Telescope, and obtained from the Hubble Legacy Archive, which is a collaboration between the Space Telescope Science Institute (STScI/NASA), the Space Telescope European Coordinating Facility (ST-ECF/ESA) and the Canadian Astronomy Data Centre (CADC/NRC/CSA). 

%%%%%%%%%%%%%%%%%%%%%%%%%%%%%%%%%%%%%%%%%%%%%%%%%%
\section*{Data Availability}

%The inclusion of a Data Availability Statement is a requirement for articles published in MNRAS. Data Availability Statements provide a standardised format for readers to understand the availability of data underlying the research results described in the article. The statement may refer to original data generated in the course of the study or to third-party data analysed in the article. The statement should describe and provide means of access, where possible, by linking to the data or providing the required accession numbers for the relevant databases or DOIs.
The spectroscopic data were obtained during ESO programmes, 171.D-0237171.0237, 182.D-0222 and 096.D-0825. Data products are available from the ESO Science Archive Portal (\url{https://archive.eso.org/scienceportal/home}). Spectra are available at \url{http://www.roe.ac.uk/~cje/tarantula/spectra, https://star.pst.qub.ac.uk/~sjs/flames/}, or on request from the authors.



\bibliographystyle{mnras}
\bibliography{Bibtex/literature.bib} % if your bibtex file is called example.bib

%\begin{thebibliography}{99}
%\end{thebibliography}

%%%%%%%%%%%%%%%%%%%%%%%%%%%%%%%%%%%%%%%%%%%%%%%%%%

%%%%%%%%%%%%%%%%% APPENDICES %%%%%%%%%%%%%%%%%%%%%

%%%%%%%%%%%%%%%%%%%%%%%%%%%%%%%%%%%%%%%%%%%%%%%%%%

% Don't change these lines
\bsp	% typesetting comment
\label{lastpage}
\end{document}